\documentclass[preprint]{aastex}
\begin{document}

\title{Optically Identified BL Lacertae Objects from the Sloan Digital Sky Survey}

\author{Matthew J. Collinge\altaffilmark{1}, Michael A. Strauss\altaffilmark{1}, 
Patrick B. Hall\altaffilmark{2,1}, \v{Z}eljko Ivezi\'c\altaffilmark{3,1}, 
Jeffrey A. Munn\altaffilmark{4}, 
David J. Schlegel\altaffilmark{5,1}, Nadia L. Zakamska\altaffilmark{1}, 
Scott F. Anderson\altaffilmark{3}, Hugh C. Harris\altaffilmark{4}, 
Gordon T. Richards\altaffilmark{1}, Donald P. Schneider\altaffilmark{6}, 
Wolfgang Voges\altaffilmark{7}, Donald G. York\altaffilmark{8}, 
Bruce Margon\altaffilmark{9}, J. Brinkmann\altaffilmark{10}}
\altaffiltext{1}{Princeton University Observatory, Princeton, New Jersey 08544 USA (email: collinge, strauss, nadia, gtr@astro.princeton.edu)}
\altaffiltext{2}{Department of Physics and Astronomy, York University, 4700 Keele Street, Toronto, ON, M3J 1P3, Canada (email: phall@yorku.ca)}
\altaffiltext{3}{Department of Astronomy, University of Washington, Box 351580, Seattle, WA 98195 USA (email: ivezic, anderson@astro.washington.edu)}
\altaffiltext{4}{U. S. Naval Observatory, PO Box 1149, Flagstaff, AZ 86002 USA (email: jam, hch@nofs.navy.mil)}
\altaffiltext{5}{Lawrence Berkeley National Laboratory, 1 Cyclotron Road, M/S 50-4049, Berkeley, CA 94720 USA (email: schlegel@astro.princeton.edu)}
\altaffiltext{6}{Department of Astronomy and Astrophysics, 525 Davey Laboratory, Pennsylvania State University, University Park, PA 16802 USA (email: dps@astro.psu.edu)}
\altaffiltext{7}{Max-Planck-Institut f{\"u}r extraterrestrische Physik, Postfach 1312, 85741 Garching, Germany (email: whv@xray.mpe.mpg.de)}
\altaffiltext{8}{Department of Astronomy and Astrophysics/Enrico Fermi Institute, University of Chicago, 5640 South Ellis Avenue, Chicago, IL 60637 USA (email: don@oddjob.uchicago.edu)}
\altaffiltext{9}{Space Telescope Science Institute, 3700 San Martin Drive, Baltimore, MD 21218 USA (email: margon@stsci.edu)}
\altaffiltext{10}{Apache Point Observatory, PO Box 59, Sunspot, NM 88349 USA (email: brinkmann@nmsu.edu)}

%----------------------------------------------------------------------%
\begin{abstract}
%\small
We present a sample of 386 BL Lacertae (BL Lac) candidates identified from 
2860~deg$^{2}$ of the 
Sloan Digital Sky Survey (SDSS) spectroscopic database. The candidates are 
primarily selected to have quasi-featureless optical spectra and low proper 
motions as measured from SDSS and USNO-B positions; however, our ability 
to separate Galactic from extragalactic quasi-featureless objects (QFOs) 
on the basis of proper motion alone is limited by the lack of reliable 
proper motion measurements for faint objects. Fortunately, 
high proper motion QFOs, mostly DC white dwarfs, 
populate a well defined region of color space, approximately corresponding 
to blackbodies with temperatures in the range 7000--12000~K. 
QFOs with measurable redshifts or X-ray or radio counterparts 
(i.e., evidence of an extragalactic/AGN nature) loosely 
follow a track in color space that corresponds to power-law continua 
plus host galaxy starlight, with typical power-law slopes in the range 
$1<\alpha_{\mathrm{opt}}<2$ ($f_{\nu}\propto \nu^{-\alpha}$).
Based largely on this remarkably clean color separation, we subdivide 
the sample into 240 probable candidates and 146 additional less 
probable (likely stellar) candidates. 
The probable BL Lac candidates have multi-wavelength 
properties consistent with the range of previously known BL Lacs, 
with an apparent preponderance of objects with synchrotron peaks at 
relatively high energies (HBL/XBL-type). The majority of the 
154 objects with measurable redshifts have z~$<1$, with a median of 0.45; 
there are also a handful of high-redshift objects extending up to z~$=5.03$. 
We identify a small number of potential radio-quiet 
BL Lac candidates, although more sensitive radio observations are needed 
to confirm their radio-quiet nature.
\end{abstract}


%----------------------------------------------------------------------%

\section {Introduction}

Among the rarest and most extreme of the various observational classes 
of Active Galactic Nuclei (AGN) are the BL Lacertae objects (BL Lacs). 
The defining traits of BL Lacs generally include:
a broad, rapidly variable, non-thermal spectral 
energy distribution (SED) with significant power from radio wavelengths to 
X-rays and $\gamma$-rays; lack of the strong optical emission lines that 
typify other AGN; high and variable polarization at 
optical and radio wavelengths; and compact, core-dominated radio 
morphology \citep{kollgaard94}. All of the above properties can be 
reproduced naturally by a model in which the dominant source of the 
observed emission is a relativistic jet of plasma directed within a 
small angle to the line of sight (e.g., \citealt{blandford78}). This 
picture is particularly appealing because there is a known class of 
objects that plausibly consists of physically identical systems in which 
the jet is directed at a more substantial angle to the line of sight,
namely the Fanaroff-Riley class {\sc I} (FR{\sc I}; low-power; 
\citealt{fanaroff74}) radio galaxies. The beamed counterparts of 
more powerful FR{\sc II} radio galaxies are generally believed to 
be flat-spectrum radio quasars (FSRQs), although FSRQs have emission 
line properties more like those of typical broad-line AGN. 
BL Lacs (at lower typical 
luminosity) and FSRQs (at higher luminosity) are often grouped 
together under the designation of blazars. This scenario is favored in unification 
schemes for radio-loud AGN \citep{urry95}.

Traditionally, BL Lacs have been discovered from observations at either 
radio (e.g., \citealt{stickel91}) or X-ray (e.g., 
\citealt{stocke91, schachter93, perlman96}) 
frequencies. Such radio and X-ray selected samples of 
BL Lacs are known to be composed of objects with systematically 
different properties, such as the peak frequencies of the 
spectral energy distributions (SEDs), as discussed by \citet{padovani95}. 
Those authors coined the terms `low-energy cutoff BL Lacs' 
(LBLs\footnote{In more recent context the word ``cutoff'', which refers to 
a break in the energy distribution of the relativistic electrons responsible 
for the synchrotron emission, is often replaced with ``peaked'', taken to 
refer to the energy corresponding to the synchrotron peak.}) and 
`high-energy cutoff BL Lacs' (HBLs) to distinguish the types of objects 
predominantly found in radio and X-ray selected samples, respectively; the 
obvious acronyms RBL and XBL are often used to make the same distinction. 
It has been proposed that blazar properties follow a continuous 
sequence from BL Lacs into the regime of FSRQs; 
in this scenario an object's observed characteristics 
are determined by the angle between the jet axis and the line of sight, the overall 
jet power, and possibly another parameter \citep{fossati98, ghisellini98}.

Large optical surveys, such as the Sloan Digital Sky Survey (SDSS; 
\citealt{sdsstechsumm}) and 
the Two-degree Field QSO Redshift Survey (2QZ; \citealt{boyle00}), 
have the potential to reveal new 
populations of BL Lacs with different properties from the well known 
radio or X-ray selected objects. 
\citet{londish02} presented a sample of several dozen objects from the 2QZ 
selected on the basis of having quasi-featureless optical spectra 
resembling those of BL Lacs; however, some of these objects were found to 
be significantly fainter in the radio, relative to the optical, 
than classical BL Lacs. 
At least one of these latter objects has been confirmed 
to be extragalactic via a redshift measurement obtained in a follow-up 
observation \citep{londish04}; along with a small number from other 
searches \citep{fan99,anderson01}, these objects are potential 
radio-quiet BL Lacs or weak-lined radio-quiet quasars. Whether or how such 
objects fit into the blazar sequence or more generally into the AGN 
unification picture are open questions.

In the present work, we describe a sample of 386 optically-identified 
BL Lac candidates compiled from 2860~deg$^2$ of the SDSS spectroscopic survey. 
As in \citet{londish02}, these objects are selected to have quasi-featureless 
optical spectra and are identified solely on the basis of their optical 
properties and lack of significant proper motions. 
We discuss the degree to which the sample remains contaminated by 
featureless stars, either lacking reliable proper motion measurements, or 
simply having proper motions below the detection threshold.
We subdivide the sample into 240 ``probable'' and 146 ``possible'' candidates 
based on all available information. We search the sample for exceptional 
objects, such as radio-quiet BL Lacs (or lineless radio-quiet quasars), and 
we explore how our sample fits into the overall understanding of the blazar class. 
Our probable BL Lac candidates are not a complete sample, but are more 
numerous than any previous sample of BL Lacs of which 
we are aware, and represent a significant addition to the $\sim 400$ 
previously known BL Lacs.

In \S\ref{sec:sdssdata} we describe the SDSS observations and data 
processing. In \S\ref{sec:selection} we discuss the selection of 
quasi-featureless spectra and the use of proper motion to reject 
weak-featured stellar objects. 
In \S\ref{sec:cross-corr} we cross-correlate our sample with 
the ROSAT All-Sky Survey (RASS) X-ray source catalogs and with the 
Faint Images of the Radio Sky at Twenty~cm (FIRST) and NRAO/VLA Sky Survey 
(NVSS) radio catalogs. In \S\ref{sec:multiwavelength} we analyze the 
redshift distribution, optical colors, and multi-wavelength properties 
of our BL Lac candidates. In \S\ref{sec:future} we lay out some prospects 
for follow-up studies, and in \S\ref{sec:conclusion} we summarize our 
primary results.

%----------------------------------------------------------------------%
\section {SDSS observations and data pipeline}
\label{sec:sdssdata}

The Sloan Digital Sky Survey  is a multi-institutional 
effort to obtain high-quality 5-band (SDSS $ugriz$; \citealt{ugrizsystem,
apophotomon,ugrizstandard}) photometric imaging of a large 
fraction ($\approx 10000$~deg$^2$) of the sky, principally the northern 
Galactic cap and a stripe along the celestial equator, with spectroscopic 
follow-up of interesting samples of objects, primarily galaxies ($\sim 10^6$) 
and quasars ($\sim 10^5$). 
The SDSS uses a dedicated 2.5~m telescope at Apache Point Observatory with 
a $2.5^{\circ}$ 30-CCD mosaic camera \citep{sdsscamera} and two 
multi-object spectrographs with 320 $3^{\prime\prime}$ fibers each 
(e.g., \citealt{sdssedr}). Imaging data are processed 
through automated software pipelines \citep{sdssphoto,
sdssastromcal}, in which the data are 
photometrically and astrometrically calibrated and objects are detected, 
measured, and classified. The photometric calibration is typically better than 2\% 
\citep{ivezic04}, and the astrometric calibration is better than 
$0\farcs 1$ root-mean-square (rms) per coordinate. 

\subsection{Targeting for spectroscopy}
\label{sec:target}

Once a region of sky has been imaged, subsets of the objects detected 
are then selected for spectroscopic observations on the basis of their 
photometric and other properties.
Resolved objects are targeted to $r=17.77$ for the main 
galaxy sample \citep{sdssgaltarget}, and luminous red galaxies are targeted based on
their colors to $r=19.5$ \citep{sdsslrgtarget}. Quasar candidates are targeted 
as outliers from the stellar locus or as point-like FIRST sources to $i=19.1$ 
for the low-redshift sample; objects very red in $u-g$, $g-r$ or 
$r-i$ (and having colors consistent with Ly$\alpha$ absorption cutting into 
the observed frame optical flux) 
are targeted to $i=20.2$ for the high-redshift quasar sample \citep{sdssqsotarget}.
Ultraviolet-excess (UVX; dereddened $u-g<0.6$) objects with non-stellar 
colors are explicitly included in the low-redshift sample. 
Objects can also be targeted by a variety of serendipity 
algorithms to $g=20.5$ or $i=20.5$. 
These selection routines cover regions of color space 
outside the main stellar locus but not included by quasar target selection; 
they also include objects with radio counterparts detected by the FIRST survey 
\citep{becker95,ivezic02} or X-ray counterparts from the ROSAT All Sky Survey 
(RASS; \citealt{voges99,anderson03}). 
Spectroscopic fibers are allocated by a tiling algorithm \citep{blanton03}, 
which gives top priority 
to the main quasar and galaxy samples (along with fibers needed for 
calibration purposes). 

\subsection{Processing of spectra}
\label{sec:spec_process}

Raw spectral data are reduced and calibrated (e.g., \citealt{sdssedr}), 
and individual spectra, which 
cover the wavelength range 3800--9200~\AA\ with an approximate resolution 
of $\lambda/\Delta\lambda\approx 1800$, are then 
automatically processed to classify them (star, galaxy, or quasar) and 
determine redshifts.

In our analysis, we use the output of the 
classification/redshift-finding routine ``SpecBS'' (written by D. Schlegel).
This algorithm classifies spectra by performing separate 
$\chi^2$-fits to the spectra using linear combinations of different models 
--- sets of galaxy or quasar eigenspectra or stellar templates --- 
plus 2nd or 3rd order polynomials, with redshift/radial velocity 
as a free parameter. An object is classified as a star, galaxy or 
quasar, according to which type of model produces the best $\chi^2$ value. 
Redshift warning flags are set if there are multiple models that provide 
nearly equivalent $\chi^2$ values or if other problems are present, such 
as limited wavelength coverage due to lost data, unreasonable behavior like 
negative templates, or large numbers of statistically anomalous data points.

The polynomial component, which may be positive or negative, 
is included to account for (generally low-level) 
variations of individual objects from the templates and to compensate for 
contamination (e.g., star-galaxy blends) or flux-calibration problems 
(cf.\ discussion in \citealt{abazajian04}), 
as well as to assist in fitting the spectra of unusual 
(e.g., heavily reddened) objects. Objects with the quasi-featureless spectra 
we seek will generally require 
strong polynomial components in order to achieve adequate fits, as discussed 
in the next section. In the most extreme cases, the polynomial components can 
completely dominate the fits; for such objects the spectral classifications 
and redshifts are clearly meaningless.

%----------------------------------------------------------------------%
\section{Sample selection}
\label{sec:selection}

In order to avoid the selection biases that enter 
into the construction of radio and X-ray-selected  
samples of BL Lacs to the greatest degree possible, we confine our 
selection criteria strictly to the optical properties of objects in the SDSS 
spectroscopic database. Of course, objects enter into 
that data set via a variety of target selection algorithms as described in 
the previous section, some of which are based on X-ray or radio emission. 
However, by keeping track of the reasons for which each object was 
targeted for spectroscopy, we can account after the fact  
for selection biases introduced by these factors (see discussion 
in \S\ref{sec:sdss_selection_effects}). 

Most of the sky coverage of SDSS is imaged in one epoch only. 
Spectroscopic observations of a given field typically lag behind 
imaging by anywhere from approximately one month to one year. 
A detailed study of the variability of spectroscopically-confirmed quasars 
using the photometric+spectroscopic observations is presented by 
\citet{vandenberk04}. For our purposes however, (usually) having only two 
epochs of observation in total for a given spectroscopic object precludes 
complete selection on the basis of variability. 
Another optical property that could be used to identify BL Lacs would be 
polarization (e.g., \citealt{jannuzi93}), but this is not among the 
capabilities of the SDSS.

Therefore, the spectroscopic phase of our BL Lac candidate selection 
is primarily based on a 
single property: lack of strong features (i.e., emission or 
absorption lines) in the optical spectra. While this approach is 
clearly biased toward objects that are dominated by the continuous emission 
from the jet at optical wavelengths (and subject to the effects of SDSS 
spectroscopic target selection), this bias should differ from 
those that affect classical radio or X-ray selected samples and 
therefore may be sensitive to new populations of BL Lacs missed 
by previous surveys. This is the same approach used by \citet{londish02}.

The parent sample for our search is the SDSS spectroscopic database 
as of mid-July 2002, comprising over 345,000 individual spectra 
and covering about 2860 square degrees of sky [roughly the area 
covered by Data Release Two (DR2); \citealt{abazajian04}]. The initial selection 
was performed using a somewhat older version of the spectroscopic data 
reduction pipeline, approximately corresponding to Data Release One 
(DR1; \citealt{abazajian03}). In that version there were (known) 
spectrophotometric problems, especially in the 3800-4400~\AA\ range, 
such as artificial structure resembling emission or absorption features; 
we employed caution in our search so as not to mistake these and 
other data reduction artifacts for real spectral structure. 
These problems have been corrected to a large degree in DR2 and 
later reductions (cf.\ \citealt{abazajian04}). As previously stated, 
we use the SpecBS spectroscopic classifications and spectral fits 
(contemporaneous with DR2); all other data presented 
in this paper (e.g., photometry, astrometry, etc.)\ are from the official 
DR2 reductions, unless stated to the contrary.

%----------------------------------------------------------------------%
\subsection {Selection of quasi-featureless spectra}
\label{sec:selection_featureless}

The goal of our search is to extract all high-quality SDSS spectra that 
are consistent with the featureless or quasi-featureless spectrum of a 
BL Lac object, without regard to the continuum slope. Depending on intrinsic 
luminosity, redshift, and the details of AGN physics, these objects may 
show (a) host galaxy signatures such as stellar absorption features 
(e.g., \ion{Ca}{2}~H\&K) or emission lines from associated \ion{H}{2} 
regions (in which case there is a good chance that the image of the 
object will also show spatially resolved structure), 
(b) weak broad or narrow emission lines associated with AGN activity, 
(c) metal absorption lines or Ly$\alpha$ forest absorption from 
intervening material along the line of sight, or 
(d) no features whatsoever, to the limit of the SDSS spectral sensitivity 
(typically better than a few \AA\ equivalent width for narrow features, 
for the objects that 
satisfy our signal-to-noise criteria; see \S\ref{sec:selection_sn}). 
Especially for the truly featureless 
objects, we reiterate that the spectral classifications and redshifts 
obtained from the spectroscopic pipeline will be meaningless.

A complication inherent in this approach is that BL Lacs are not the only 
type of objects known to have featureless optical spectra. In particular, 
white dwarfs (WDs) of class DC share this property 
(e.g., \citealt{wesemael93,londish02}).
This is not an issue for objects that fall into classes (a)--(c) above. 
To distinguish true BL Lacs of class (d) from DC WDs 
without resorting to multi-wavelength properties, the best available test is 
to use proper motion (see \S\ref{sec:selection_contaminants} below). 
In what follows, we shall refer to objects that satisfy our spectral 
selection criteria as quasi-featureless objects (QFOs); we shall reserve 
the term ``BL Lac candidate'' for objects that satisfy the additional proper 
motion requirement described in \S\ref{sec:selection_contaminants}.

\subsubsection{Signal-to-noise requirement}
\label{sec:selection_sn}

In order to restrict our analysis to objects for which reliable spectral 
classifications are possible, we first impose a signal-to-noise ratio ($S/N$) 
requirement. We analyze the spectral flux density 
(erg~cm$^{-2}$~s$^{-1}$~\AA$^{-1}$) in three 500~\AA-wide regions 
centered at 4750, 6250, and 7750~\AA\ in the observed frame; 
these are chosen to be regions in which 
we find SDSS spectra to be of generally high quality, with minimal sky 
contamination and calibration problems, and which fall within the $g$, $r$ and $i$ 
filters respectively. We define the signal $S$ in each of 
these regions to be $S=\sum_i (f_{\lambda,i}/\sigma_i^2)/\sum_i (1/\sigma_i^2)$, 
where $f_{\lambda,i}$ and $\sigma_i$ are the flux density and 
estimated uncertainty in spectral element $i$. The error $N$ on this 
quantity is simply $N=[\sum_i 1/\sigma_i^2]^{-1/2}$; 
therefore the signal-to-noise ratio for each region is 
$S/N=(\sum_i f_{\lambda,i}/\sigma_i^2)/(\sum_i 1/\sigma_i^2)^{1/2}$. 
We keep only objects for which $S/N>100$ in one or more of these spectral regions. 
This corresponds to fiber magnitudes of roughly $g<20.5$, $r<20.3$ 
or $i<19.6$ for a typical spectroscopic plate.

\subsubsection{Objects classified as stars or galaxies}
\label{sec:selection_stargal}

Stars and galaxies together account for the majority (89\%) of all SDSS 
spectroscopic objects. Most of these are 
well fit by stellar templates or combinations of galaxy eigenspectra, respectively, 
without the need for significant polynomial components in the fits 
(see \S\ref{sec:spec_process}). 
This is true only for objects which have characteristic stellar 
or galactic spectral energy distributions (SEDs) 
and absorption features of typical strength; such objects clearly do not 
qualify as quasi-featureless. In order to separate such objects from 
potential QFOs, we examine the 
relative importance of the polynomial component in the fit in two spectral 
regions: 3800--4200~\AA\ and 8800--9200~\AA\ (in the observed frame; 
near the blue and red ends respectively 
of the SDSS spectral coverage). We define the ratio 
$R\equiv P/S$, where $P$ is the average value of the 
polynomial in a region and $S$ is the spectral signal as defined in 
the previous subsection. If $S/N<25$ in both regions, the values of $R$ are not 
very meaningful and the test cannot be applied; such objects are rejected from 
further consideration. 
(This is similar to the $S/N$ requirement described in the previous subsection but 
much less stringent.) We reject objects classified by SpecBS as stars or galaxies 
for which $R_{\mathrm{blue}}<0.6$ and $R_{\mathrm{red}}<0.5$; the blue cutoff 
is more stringent due to the large number of absorption features in this 
wavelength region for many stellar types and the presence of the 
\ion{Ca}{2} H\&K break in galaxy spectra. 
This step eliminates approximately 80\% of stars and 96\% of galaxies. 

We do not reject objects classified as quasars based on the values of $R$. 
Due to the range of quasar emission line strengths and the power-law 
nature of quasar continuum emission, some quasi-featureless objects may be well fit 
by combinations of quasar eigenspectra (through the observed wavelength range) 
without the need for a strong polynomial component.

\begin{figure}
\epsscale{1.0}
\plotone{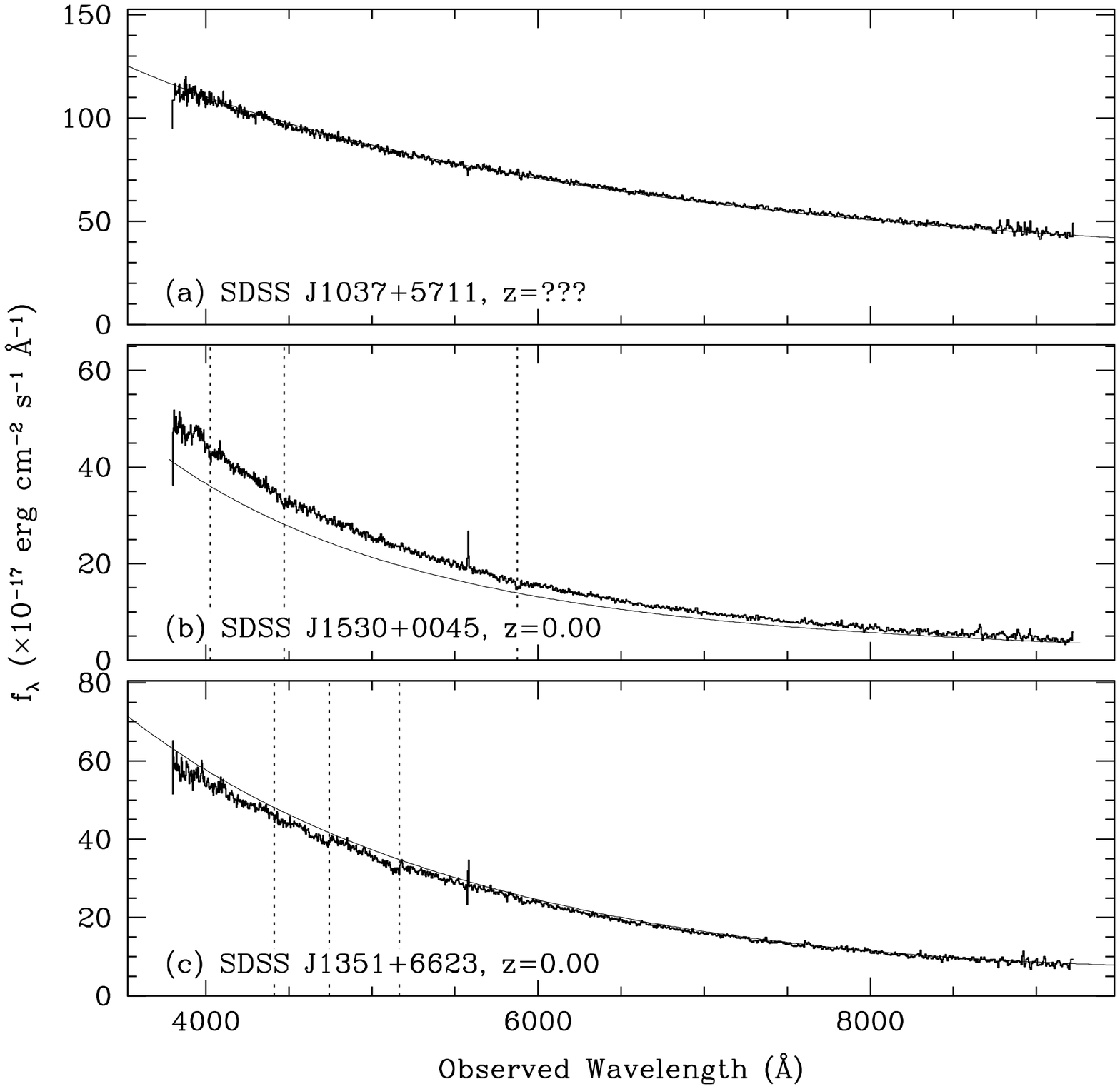}
\figcaption{
Example weak-featured SDSS spectra. The smooth solid curves show the polynomial 
components from the pipeline spectral fits. 
The feature at 5577~\AA, most prominent in panels (b) and (c), is 
a residual from sky subtraction. Two out of the three objects 
were mis-classified by the spectral pipeline and assigned erroneous redshifts; 
in all three cases redshift warning flags were set. 
The dotted vertical lines mark the 
positions of apparent absorption features. Panel (a): no features are detected; 
this object survives the stellar absorption line cut described in 
\S\ref{sec:selection_stargal} and enters the sample of QFOs.
Panel (b): \ion{He}{1} $\lambda\lambda$ 4026, 4471, 5876~\AA\ are detected 
in absorption; thus this object is probably a DB WD.
Panel (c): shallow Swan bands of the C$_2$ molecule are detected (dotted lines 
mark their approximate edges); this object is a DQ WD. Objects (b) 
and (c) are rejected as QFOs. 
\label{fig:featureless_examples}}
\end{figure}

Objects classified as stars that survive the previous cut are then subjected 
to a test for absorption features appropriate to the SpecBS-fit stellar type. 
We test stars classified as O, B, or A for absorption lines from \ion{He}{2}, 
\ion{He}{1}, and the H-Balmer series. We test F and G stars for the 
H-Balmer series and lines from \ion{Ca}{2}, \ion{Mg}{1}, and \ion{Na}{1}. We 
test K stars for \ion{Ca}{1}, \ion{Ca}{2}, \ion{Mg}{1}, and \ion{Na}{1} lines.
We test M stars for \ion{Ca}{2} and \ion{Na}{1} lines and TiO molecular absorption 
bands. We measure the significance of all absorption features by comparing 
the spectral flux in pre-defined continuum regions on either side of the absorption 
to the spectral flux within the absorption feature.
Objects are rejected if two or more absorption features 
are detected with $\geq 3\sigma$ significance; we require the detection of 
at least two features to decrease the chance of inadvertently rejecting 
extragalactic objects showing absorption lines from intervening material. 
Examples of objects with weak stellar features or featureless spectra are 
displayed in Figure~\ref{fig:featureless_examples}.

\begin{figure}
\epsscale{1.0}
\plotone{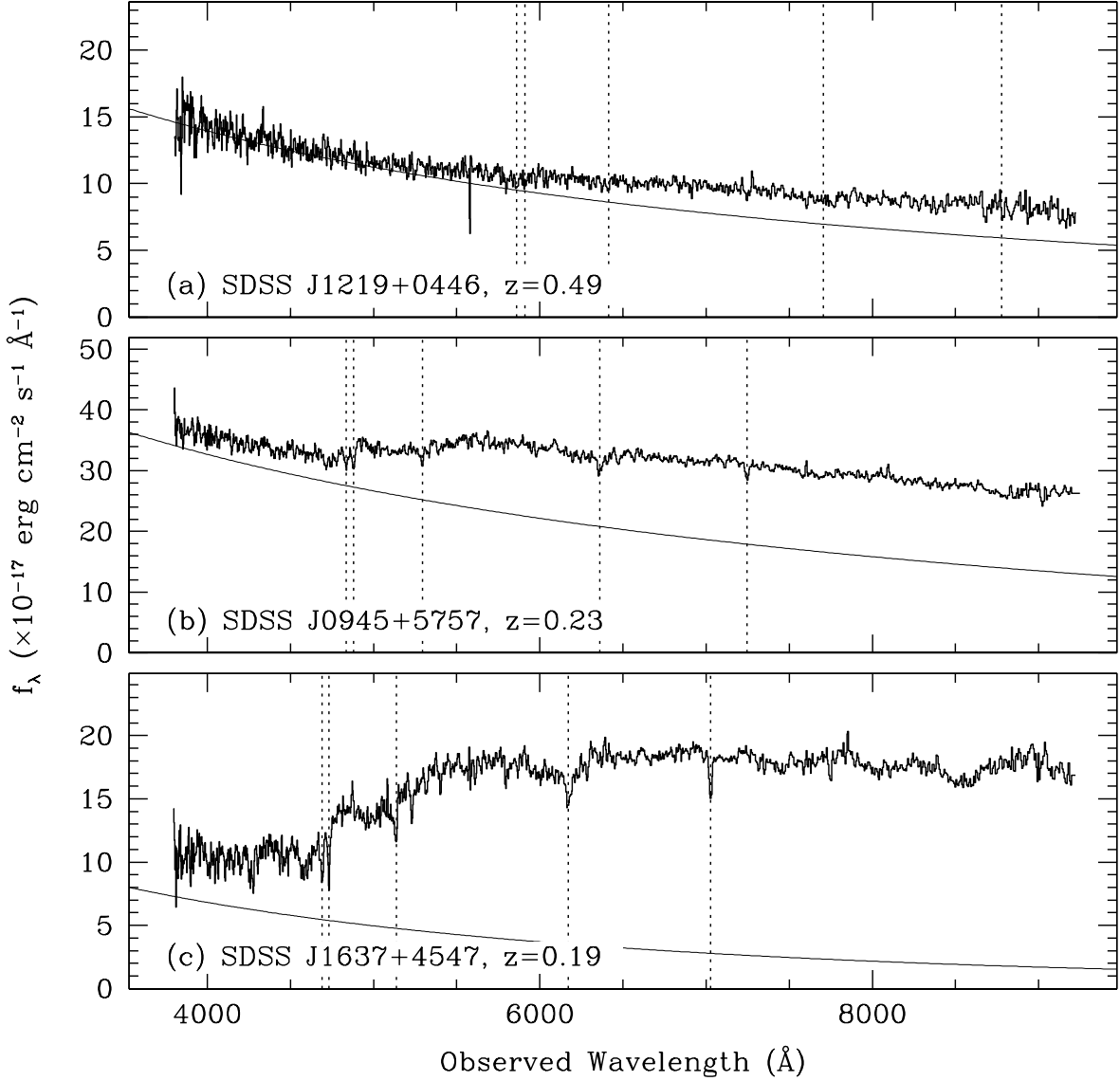}
\figcaption{
Example SDSS spectra of QFOs with noticeable host galaxy 
features of varying strengths and secure redshift measurements. The format is the 
same as Figure~\ref{fig:featureless_examples}. The dotted 
lines mark the positions of stellar absorption lines at the appropriate redshifts. 
Objects (a) and (b) are unambiguous BL Lac candidates. Object (c) 
represents a marginal case for a BL Lac: it has a \ion{Ca}{2} H\&K break 
strength of 0.36, larger than the classical cutoff of 0.25 but within our 
acceptable range of $<0.4$. All three objects enter the sample of QFOs.
\label{fig:galaxy_examples}}
\end{figure}

We test objects classified as galaxies for the strength of the \ion{Ca}{2} 
H\&K break (at the pipeline redshift) if it is present in the spectra and 
if no redshift warning flags have been set. The \ion{Ca}{2} 
H\&K break strength is defined as
\begin{equation}
\label{eq:cahk}
C=0.14+0.86(f_{\lambda,+}-f_{\lambda,-})/f_{\lambda,+},
\end{equation} 
where $f_{\lambda,-}$ and $f_{\lambda,+}$ are the average flux densities 
in the rest-frame wavelength regions 3750--3950~\AA\ and 
4050--4250~\AA\ respectively (e.g., \citealt{landt02}). (The form of this 
equation derives from the fact that it is historically measured in $f_{\nu}$ 
as in \citealt{bruzual83}, rather than $f_{\lambda}$.)
We reject objects for which $C>0.4$ (as in \citealt{marcha96}) 
and $|C/\delta C|>3$ where $\delta C$ is the uncertainty on $C$. 
\citet{stocke91} adopted a more conservative cut of 0.25, 
but we err on the liberal side 
in order not to reject any intrinsically red BL Lacs with 
significant host galaxy contamination. Several examples of objects with 
a range of host galaxy contributions that survive the polynomial and 
\ion{Ca}{2} H\&K cuts are given in Figure~\ref{fig:galaxy_examples}.

\subsubsection{Objects classified as quasars and other emission line objects}
\label{sec:selection_emissionline}

For all objects classified as quasars, and for objects classified as stars or 
galaxies that survive the cuts described in the previous subsection, we 
test for the presence of emission lines at the pipeline redshift. 
We measure the equivalent width (EW) of an emission line by comparing the 
integrated flux in a pre-defined wavelength region around the center of the line 
with a linear approximation to the continuum in that region of the spectrum. 
The continuum level is determined by examining the flux in pre-defined regions 
on either side of the line. Special care is taken to examine potential continuum 
regions for the presence of absorption lines (e.g., metal lines from intervening 
material), which could artificially lower the continuum level and thereby 
exaggerate EW measurements. We do not attempt to automatically measure the EW 
of the Ly$\alpha$ emission line due to the frequently-complex absorption 
toward shorter wavelengths. This is to protect against inadvertently rejecting any 
BL Lacs at high enough redshift to show absorption from the Ly$\alpha$ forest 
(e.g., \citealt{fan99}).

We reject all objects in which any identified emission line has a 
rest-frame equivalent width (RFEW) greater than 5~\AA\ 
at 1-$\sigma$ (i.e., the 1-$\sigma$ error on RFEW 
does not overlap 5~\AA). This criterion is in accordance with the 
classical definition of a BL Lac, but is rather strict by modern 
standards (e.g., \citealt{marcha96}) in that it may cause us to reject 
objects with featureless AGN continua plus emission lines from host-galaxy 
star formation. However, we employ it in order not to be overwhelmed by 
galaxies with moderate star formation rates, which exist in the 
SDSS spectroscopic data in large numbers. 

A line identification is considered secure only if 
other spectral features are present at the same redshift; thus a line 
identification implies that a redshift can be determined. 
If a single emission feature appears to be present but 
a line identification cannot be unambiguously determined, we do not 
reject the object. These criteria are designed so that quasi-featureless objects 
(possibly having bumpy spectra, e.g., due to intervening absorption) will 
not be rejected no matter what redshift the pipeline assigns them. 
Example spectra of borderline emission-line objects are shown 
in Figure~\ref{fig:line_examples}. 

\subsubsection{Manual examination and frequent contaminants}

\begin{figure}
\epsscale{1.0}
\plotone{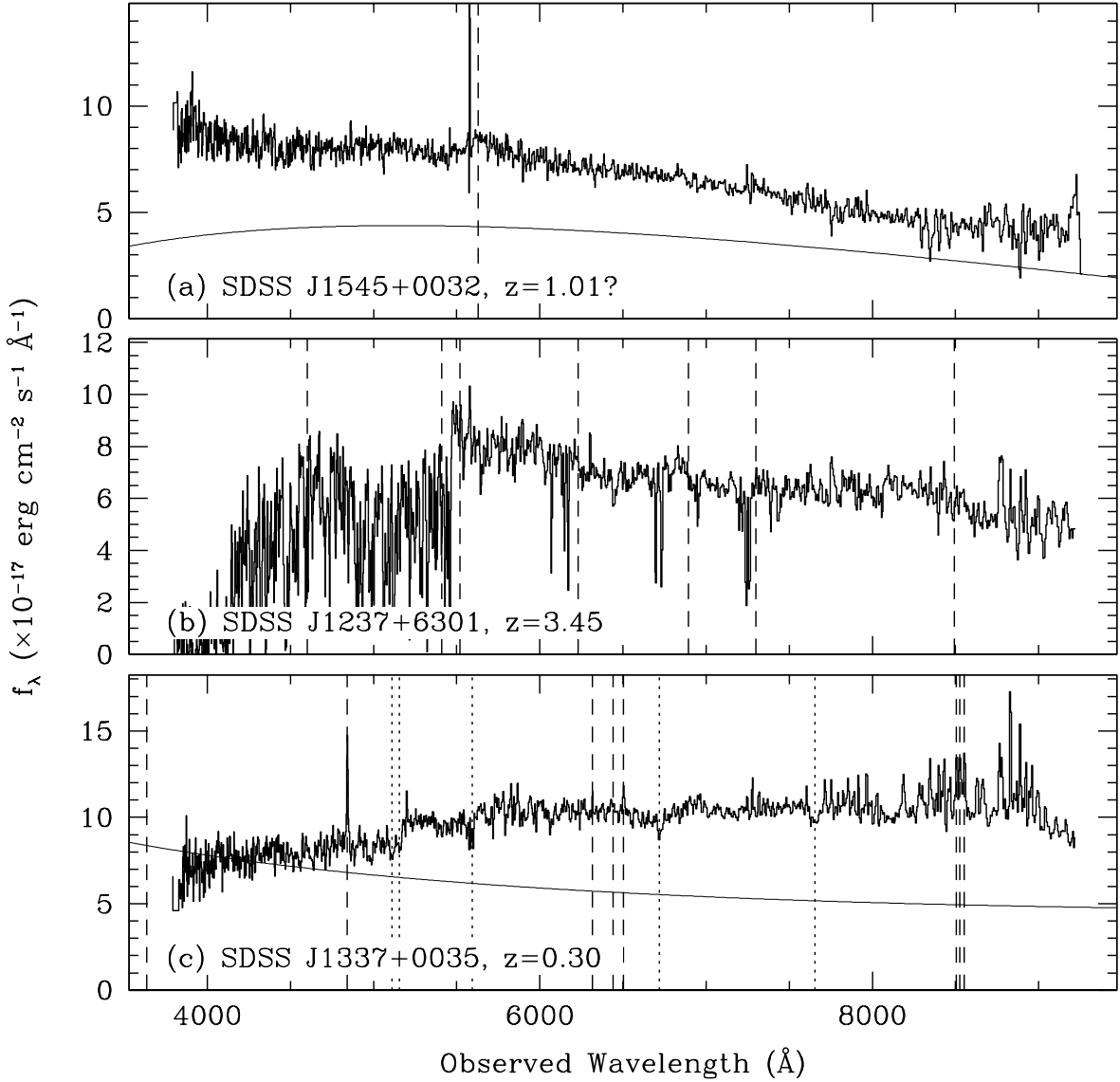}
\figcaption{
Example SDSS marginal emission-line objects, similar to 
Figs.\ \ref{fig:featureless_examples} and~\ref{fig:galaxy_examples}. 
Dotted lines again mark absorption features, and 
dashed lines mark emission features. All three spectra
show significant sky residuals at the red ends; 
many of the apparent emission features long-ward of about 7500~\AA\ cannot 
be trusted. Panel (a): the broad 
emission-line-like feature observed near 5630~\AA\ may be \ion{Mg}{2} 
$\lambda$~2800~\AA; however, the identification is not secure primarily 
due to the lack of other features, so this object survives the emission 
line cut described in \S\ref{sec:selection_emissionline} and enters 
the sample of QFOs. Panel (b): the broad lines appear weak and/or absorbed in 
this object; the \ion{C}{4} $\lambda$~1549~\AA\ line (observed near 6800~\AA\ and 
possibly blue-shifted with respect to Ly$\alpha$) has RFEW close to 5~\AA, but with 
substantial uncertainty. Therefore object (b) survives as a QFO.
Panel (c): \ion{O}{2} $\lambda$~3727~\AA\ has RFEW$>$5~\AA.
Object (c) is rejected as a QFO.
\label{fig:line_examples}}
\end{figure}

The selection procedure described above is performed by an 
automated algorithm; however, the algorithm is designed to 
err on the side of caution so as not to inadvertently reject unusual and 
interesting objects. The result is that 3527 spectra 
(1.02\% of the parent sample) survive through 
the automated stage and must be manually examined. 
At this stage we stringently reject remaining emission-line objects for which 
unambiguous line identifications can be made (with $RFEW>5$\AA), 
objects with visibly-identifiable stellar features near zero redshift, 
and objects with data reduction problems 
or various forms of contamination (e.g., star-galaxy blends).
All objects in Figures~\ref{fig:featureless_examples}--\ref{fig:line_examples} 
survived until at least the manual selection stage due to the conservative 
design of the automated algorithm.

In addition to the quasi-featureless objects we seek, several distinct types 
of objects survive in significant quantities until the manual examination stage. 
Among these are quasars with relatively weak emission lines such as the object 
shown in panel (b) of Figure~\ref{fig:line_examples}. Broad absorption line (BAL) 
quasars (e.g., \citealt{hall02, reichard03}) also survive in large numbers. Their 
frequently complicated spectra render it difficult to define continuum regions 
for measuring emission line EWs; we reject objects in which BAL troughs prohibit 
emission line measurements. Another class of frequent contaminants are blue,
low-surface brightness galaxies. The spectra of these objects often show 
weak or washed-out absorption features; however, they are easily identified by 
their image properties. A final type of frequent contaminant are 
white dwarfs with weak-to-undetectable absorption features; many of these 
objects survive through the manual examination step and are 
discussed in the following subsection. Among weak-featured WDs several types 
stand out, including weak DQ (carbon) WDs (e.g., \citealt{liebert03}) and weak 
DB (helium) WDs. \citet{kleinman04} present a catalog of WDs from SDSS, including 
many with very weak spectral features or apparently featureless spectra.

%----------------------------------------------------------------------%
\subsection {Removing contamination by weak-featured white dwarfs}
\label{sec:selection_contaminants}

After the final examination of all remaining spectra, we are left 
with a sample of 580 quasi-featureless objects out of the original 
$\approx 345,000$. Of these, many are expected to be DC white dwarfs, which 
also fulfill the requirement of having featureless optical spectra. 
(A total of 124 of the 580 QFOs appear in the \citealt{kleinman04} catalog; cf.\ 
\S\ref{sec:pmcut} and~\ref{sec:sample}.)
Additionally there may be other weak-lined stellar objects and possibly 
extragalactic non-BL Lac QFOs (e.g., \citealt{fan99,londish04}), 
which would be of great interest. 

In order to separate Galactic from extragalactic objects, several tests can 
be performed. One is the determination of a redshift/radial velocity based on 
spectral features. A second property that can indicate an extragalactic nature 
is image morphology: well-resolved objects (as reported by the photometric 
pipeline) in SDSS are mostly galaxies. However, these tests are redundant in 
many cases, since QFOs in which the host galaxy 
component is strong enough to be detected in the image also generally show 
noticeable host galaxy features in the spectra. 
Furthermore, these tests work most poorly for the most featureless objects.

A superior test is a measurement 
of proper motion, as in \citet{londish02}. Galactic objects 
may have measurable proper motions, depending on kinematic properties, distance 
from the observer, astrometric precision and time baseline. Since 
DC WDs have low luminosities, those that are bright enough for SDSS spectroscopy 
should be relatively nearby, within a few hundred~pc. For reference, a transverse 
velocity of 20~km~s$^{-1}$ at a distance of 200~pc corresponds to a proper motion 
of 21~mas~yr$^{-1}$. 
On the other hand the proper motions of extragalactic objects such as 
BL Lacs should be consistent with zero.

As of DR2, proper motions for SDSS objects are determined 
by matching with the United States Naval Observatory-B1.0 
catalog (USNO-B; \citealt{monet03}). In the area covered by the SDSS, 
the USNO-B catalog is mainly derived from POSS-I and POSS-II plates, 
with time baselines in the range 20--51~yr. 
More accurate proper motions can be 
realized by using SDSS positions to recalibrate USNO-B astrometry; 
the maximum time baselines (POSS-I to SDSS) lie in the range 33--53~yr 
for objects discussed in this paper. This approach 
has been implemented by \citet{munn04} and \citet{gould04}. 
We use the \citet{munn04} catalog, extended to include SDSS objects 
beyond DR1; all but four of the 580 QFOs appear in the extended catalog 
at the time of writing.

\subsubsection{Proper motion assessment}
\label{sec:pm_assess}

Some SDSS objects do not have counterparts in USNO-B due to the difference 
in sensitivity and other factors, and mismatches occur in the proper motion 
catalog for this and other reasons. Therefore it is necessary to evaluate, 
based on available information, whether an individual proper motion measurement 
is or is not reliable. We consider proper motions only for 
objects that are spatially unresolved in the SDSS images; the \citet{munn04} 
approach uses galaxies (resolved objects) to recalibrate USNO-B astrometry, 
and thus the proper motions of resolved objects are (almost always) consistent 
with zero by construction. We adopt the following set of reliability criteria: 
(1) we require a one-to-one match between SDSS and USNO-B; (2) we require that 
the object be detected on at least four of the five USNO-B plates, including 
both POSS-I plates; (3) we require that the 
rms residuals for the proper motion fit be less than 525~mas~yr$^{-1}$ in both 
right ascension and declination; (4) we require 
that there be no neighbor with $g<22$ within 7~arcsec. Requirements (1) 
and (2) ensure that the match between SDSS and USNO-B is likely to be correct. 
A larger residual than required in (3) indicates that the fit to the proper 
motion was poor, in that at least one detection lay well off a linear fit; 
525~mas~yr$^{-1}$ is an empirical cut (more lenient than that employed by 
\citealt{munn04}, but the same as used by \citealt{kilic05}) that 
includes the full distribution of residuals seen for good fits. Requirement (4) 
is designed to avoid errors introduced by blending on the Schmidt plates that 
make up the USNO-B catalog and is discussed in detail by \citet{kilic05}.
According to these criteria, 334 out of 580 QFOs have reliably measured 
proper motions.

\begin{figure}
\epsscale{1.0}
\plotone{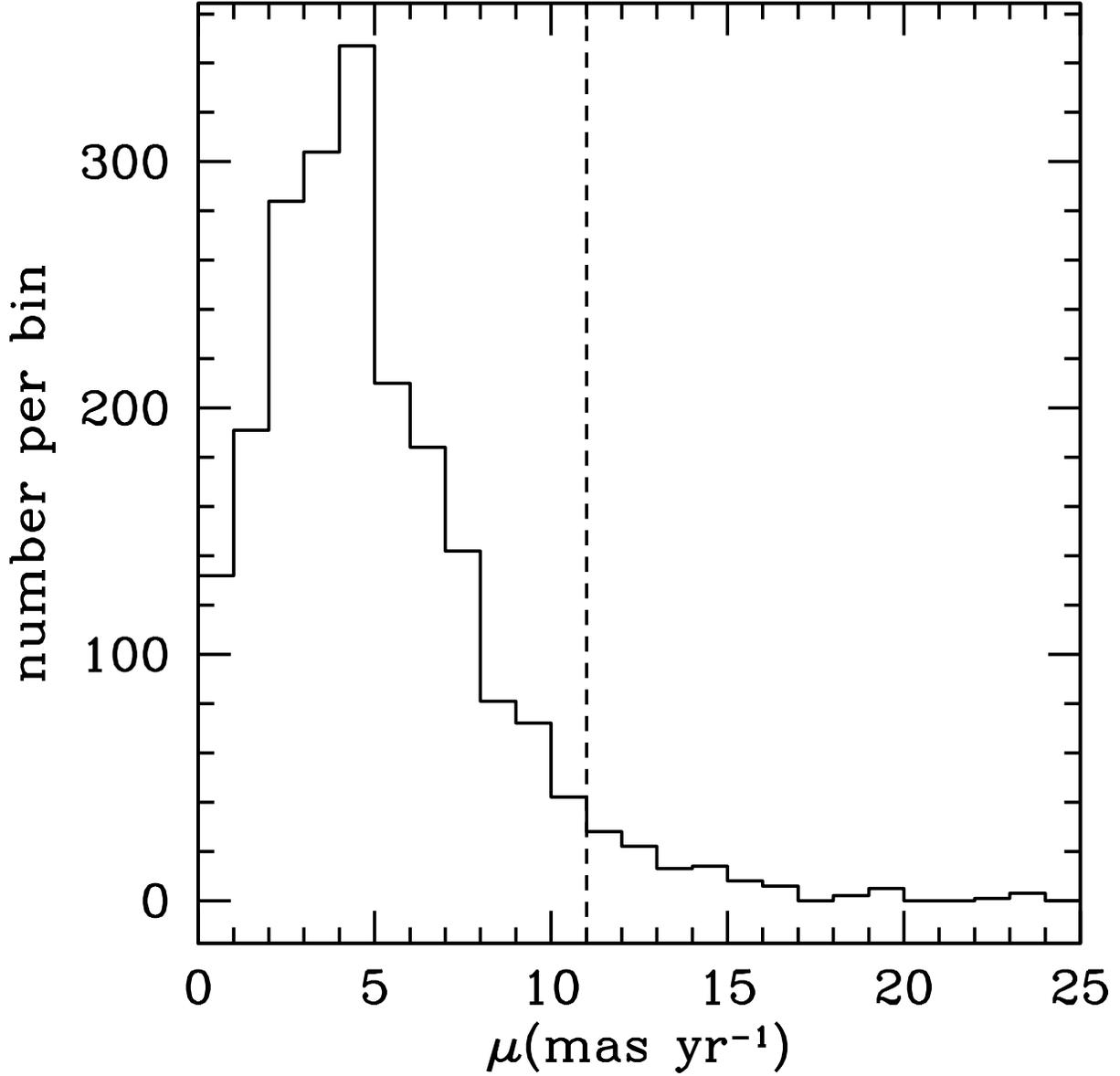}
\figcaption{
Distribution of measured proper motions for 2092 spectroscopically 
confirmed SDSS quasars. Proper motions are obtained from the \citet{munn04} 
catalog, extended to include SDSS objects not in DR1, and are required 
to satisfy reliability criteria as described in \S\ref{sec:pm_assess}.
The vertical dashed line marks $\mu=11$~mas~yr$^{-1}$; 95.0\% of the 
quasars have measured proper motions below this level.
\label{fig:pmdist}}
\end{figure}

\begin{figure}
\epsscale{1.0}
\plotone{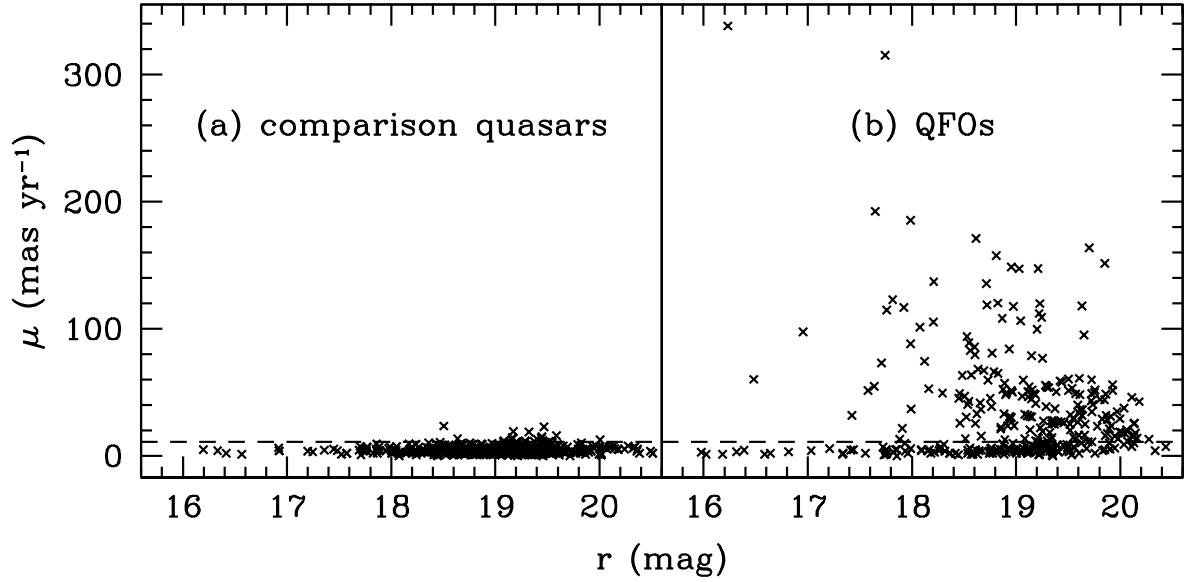}
\figcaption{
Reliably measured proper motions vs.\ $r$-band magnitude for comparison 
quasars and quasi-featureless objects. Horizontal dashed lines mark 
the 11~mas~yr$^{-1}$ level; for clarity, only 500 of the comparison quasars 
(randomly selected) are shown.
\label{fig:pmmag}}
\end{figure}

In order to confirm the utility of these ``reliable'' proper motions to 
distinguish between Galactic and extragalactic objects, we compiled a sample 
of 3202 spectroscopically confirmed SDSS quasars with secure redshifts 
(and therefore no doubt as to their extragalactic nature). Of these, 2092 
have reliably measured proper motions. Figure~\ref{fig:pmdist} shows 
the distribution of reliable quasar proper motions; 
1987/2092$=$95.0\% have measured proper motions $\mu<11$~mas~yr$^{-1}$.
This is almost identical to the \citet{munn04} result
despite the use of somewhat different reliability criteria. 
Figure~\ref{fig:pmmag} shows reliable proper motions as a function of 
$r$-band magnitude for comparison quasars and QFOs. It is evident that a 
proper motion cut of 11~mas~yr$^{-1}$ effectively separates out a large number 
of QFOs with unambiguously significant proper motions, even toward the 
faint end. Therefore, in what follows 
we shall consider proper motions $\mu\geq 11$~mas~yr$^{-1}$ to be significant, 
and proper motions $\mu<11$~mas~yr$^{-1}$ to be insignificant, unless 
otherwise stated. 

\subsubsection{Proper motions and optical colors}
\label{sec:pmcolor}

\begin{figure}
\epsscale{1.0}
\plotone{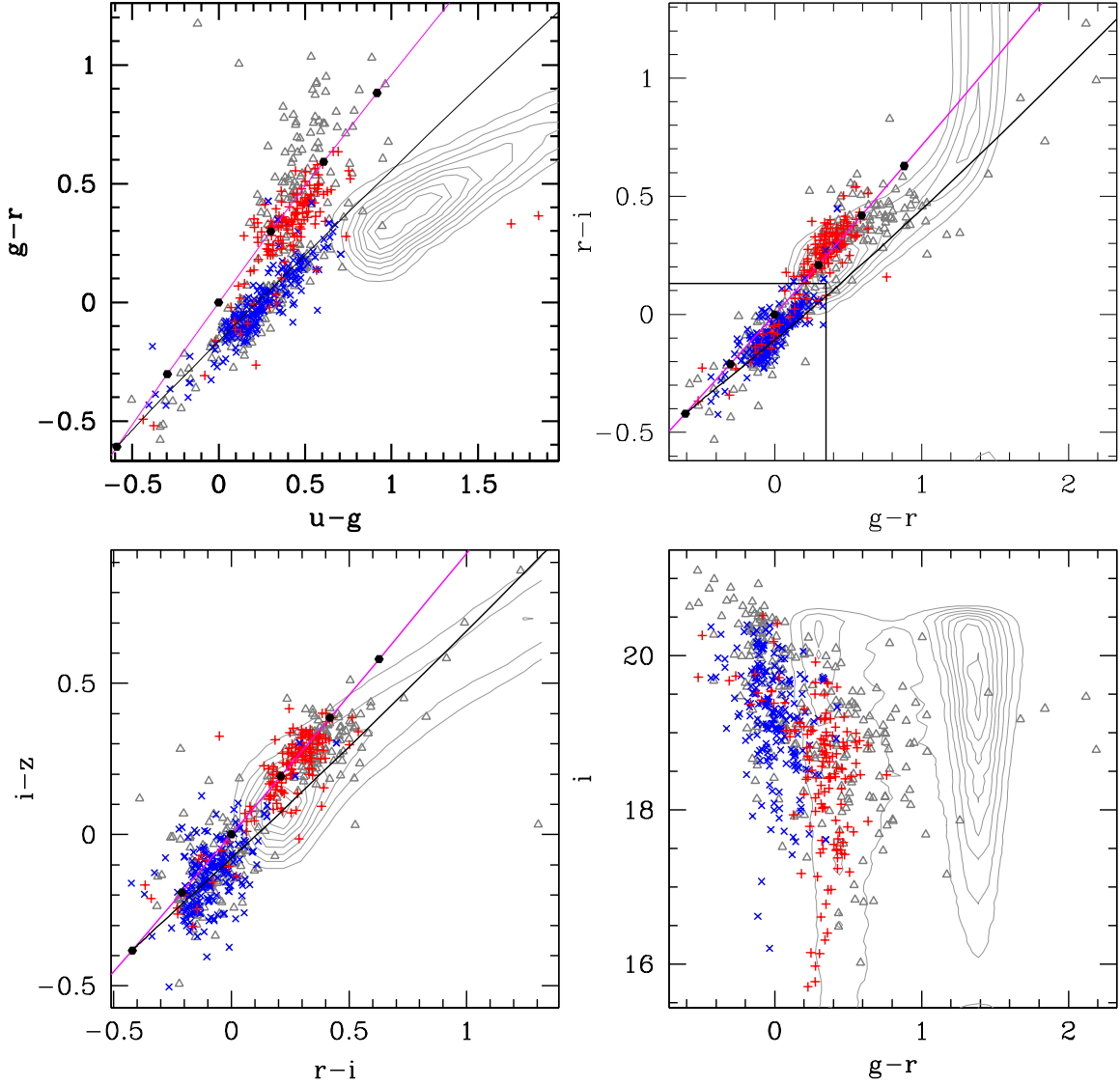}
\figcaption{\footnotesize
Color-color and color-magnitude diagrams of SDSS quasi-featureless 
objects; no correction for Galactic reddening has been applied. 
Contours represent the stellar locus (e.g., \citealt{finlator00}). 
Diagonal magenta lines in the color-color diagrams show the loci of objects with 
pure power-law spectra, black dots on these lines correspond to 
$\alpha=-2,-1,0,1,2,3$ ($f_{\nu}\propto \nu^{-\alpha}$) from lower left to 
upper right, and diagonal black lines show the loci of blackbody colors.
Red `+' signs show objects with reliably measured proper motions 
$\mu<11$~mas~yr$^{-1}$; blue `$\times$' marks are those with reliable 
proper motions $\mu\geq 11$~mas~yr$^{-1}$; grey triangles are objects that 
lack reliable proper motion measurements. The apparent correlation 
between $g-r$ color and $i$ magnitude 
is due to a combination of SDSS target selection and $S/N$ selection effects, 
as discussed in \S\ref{sec:selection_contaminants}. 
The solid-outline box in the upper right panel is the ``blue-$gri$'' region.
\label{fig:pmcolor}}
\end{figure}

Figure~\ref{fig:pmcolor} shows SDSS color-color and color-magnitude diagrams 
for quasi-featureless objects, symbolically divided according to proper motion 
at 11~mas~yr$^{-1}$. We have not corrected for the Galactic extinction/reddening, 
since many of the QFOs are expected to be DC WDs; these objects have low 
luminosities (absolute magnitudes $M_{\mathrm{B}}\sim 14$; \citealt{mccook99}), 
and thus should be relatively nearby and may not be significantly reddened.
[In Figure~\ref{fig:pmcolor} and elsewhere, we use 
point-spread function (PSF) magnitudes.] 

In addition to a handful of QFOs scattered around in color space, 
two populations of QFOs are evident in Figure~\ref{fig:pmcolor}. 
The objects with significant proper motions are, as a group, blue in 
$g-r$, $r-i$ and $i-z$, and have colors that approximately follow the 
blackbody color locus for temperatures in the range 7000--12000~K. 
There is also a large group of objects without 
significant proper motions that approximately follow the power-law color 
locus and extend toward the red; these objects are mostly concentrated 
in the range $1<\alpha_{\mathrm{opt}}<2$. (We define all slope parameters 
$\alpha$ in the sense $f_{\nu}\propto \nu^{-\alpha}$.) 
These two groups of objects are remarkably 
distinct from one another. The region defined by ($g-r<0.35$; $r-i<0.13$) --- 
the ``blue-$gri$'' region --- as delineated in the upper right panel of 
Figure~\ref{fig:pmcolor}, contains the vast majority (191/197) of objects 
with significant proper motions, while it contains only 23/137 of the 
objects without significant proper motions. 

It is plausible that the 23 blue-$gri$ QFOs with reliable proper motions 
$\mu<11$~mas~yr$^{-1}$ simply 
represent the low-proper-motion extension of the population of QFOs 
with significant proper motions. The fraction (6/120=5\%) 
of QFOs outside the blue-$gri$ region that have reliable proper motions 
$\mu\geq 11$~mas~yr$^{-1}$, on the other hand, 
is consistent with the false positive rate for 
quasars reported in the previous section. The 
bottom line is that, for objects with reliably measured proper motions, the 
blue-$gri$ region contains almost exclusively stars, whereas nearly all 
the objects outside the blue-$gri$ region have proper motions consistent 
with an extragalactic nature. We shall examine this proposition 
more carefully in the following sections.

Of the 246 QFOs without reliable 
proper motion measurements (including all QFOs with spatially resolved 
morphologies), 133 lie within the blue-$gri$ region, and 113 fall outside 
of it. Thus out of 580 QFOs, a total of 347 lie within the blue-$gri$ region, 
and 233 lie outside of it.

The anti-correlation between $g-r$ color and $i$ magnitude that is apparent at 
the blue end, in the lower right panel of Figure~\ref{fig:pmcolor}, 
is probably due to a combination of two selection effects. (1) Very blue 
quasi-featureless objects are almost exclusively stars; for the brighter objects 
weak absorption features can usually be seen in the spectra due to higher $S/N$ 
and thus the objects are rejected. Hence, there are no bright blue objects in 
the sample. (2) Many of these objects are targeted for spectroscopy 
by the blue-serendipity algorithm; for this algorithm the most stringent 
faint magnitude cutoffs are in the $u$ and $g$ bands. At a constant $u$ or 
$g$ magnitude, the bluest objects have the faintest $i$ magnitudes.

\subsubsection{Proper motion cut}
\label{sec:pmcut}

Within the blue-$gri$ region, we reject all 191 QFOs with reliable 
$\mu\geq 11$~mas~yr$^{-1}$ from further consideration as BL Lac 
candidates. While the fraction of QFOs 
outside the blue-$gri$ region that have reliable $\mu\geq 11$~mas~yr$^{-1}$ 
is consistent with the false positive rate for comparison quasars, 
a handful of objects stand out.
Out of 2092 comparison quasars with reliable proper motions, the maximum 
measured proper motion is 40.7~mas~yr$^{-1}$; three QFOs outside 
the blue-$gri$ region have measured proper motions greater than this 
value (44.3, 120.3 and 151.6~mas~yr$^{-1}$). These three objects fall 
just outside the blue-$gri$ region, all having $g-r<0.35$ but 
$0.13<r-i<0.16$. We regard their proper motions as genuine, 
and reject them as BL Lac candidates. On the other 
hand, the three QFOs outside the blue-$gri$ region that have reliable 
proper motions $11\leq \mu \leq 40.7$~mas~yr$^{-1}$ are well separated 
from the blue-$gri$ objects. They have proper motions (13.0, 14.1 and 
23.2~mas~yr$^{-1}$) that are fully consistent with the tail of the quasar 
proper motion error distribution. We regard these proper motions as dubious, 
and therefore do not reject these objects. Thus, our proper motion 
cut eliminates a total of 194 QFOs, leaving a total of 386 BL Lac 
candidates.

Since 246 of the QFOs lack reliable proper motion measurements 
(and since some DC WDs may have proper 
motions below the detection threshold), the 
proper motion cut alone cannot produce a clean sample of extragalactic 
objects. Of the 124 \citet{kleinman04} objects that qualify as QFOs 
according to our definition, 39 survive the proper motion cut 
(cf.\ \S\ref{sec:sample}); these objects all lack reliable proper motions. 

We checked the 194 QFOs rejected by the proper motion cut for other 
properties suggestive of a possible extragalactic/AGN nature. Referring 
ahead to \S\ref{sec:cross-corr}, none of these objects has a RASS counterpart
within one~arcmin, a FIRST counterpart within two~arcsec, or an NVSS counterpart 
within 10~arcsec; thus it appears that none of the QFOs rejected 
by the proper motion cut are radio or X-ray sources. Additionally, 
none of the QFOs rejected by the proper motion cut has a measurable redshift.

\subsubsection{Resolved objects and objects with redshifts}

\begin{figure}
\epsscale{1.0}
\plotone{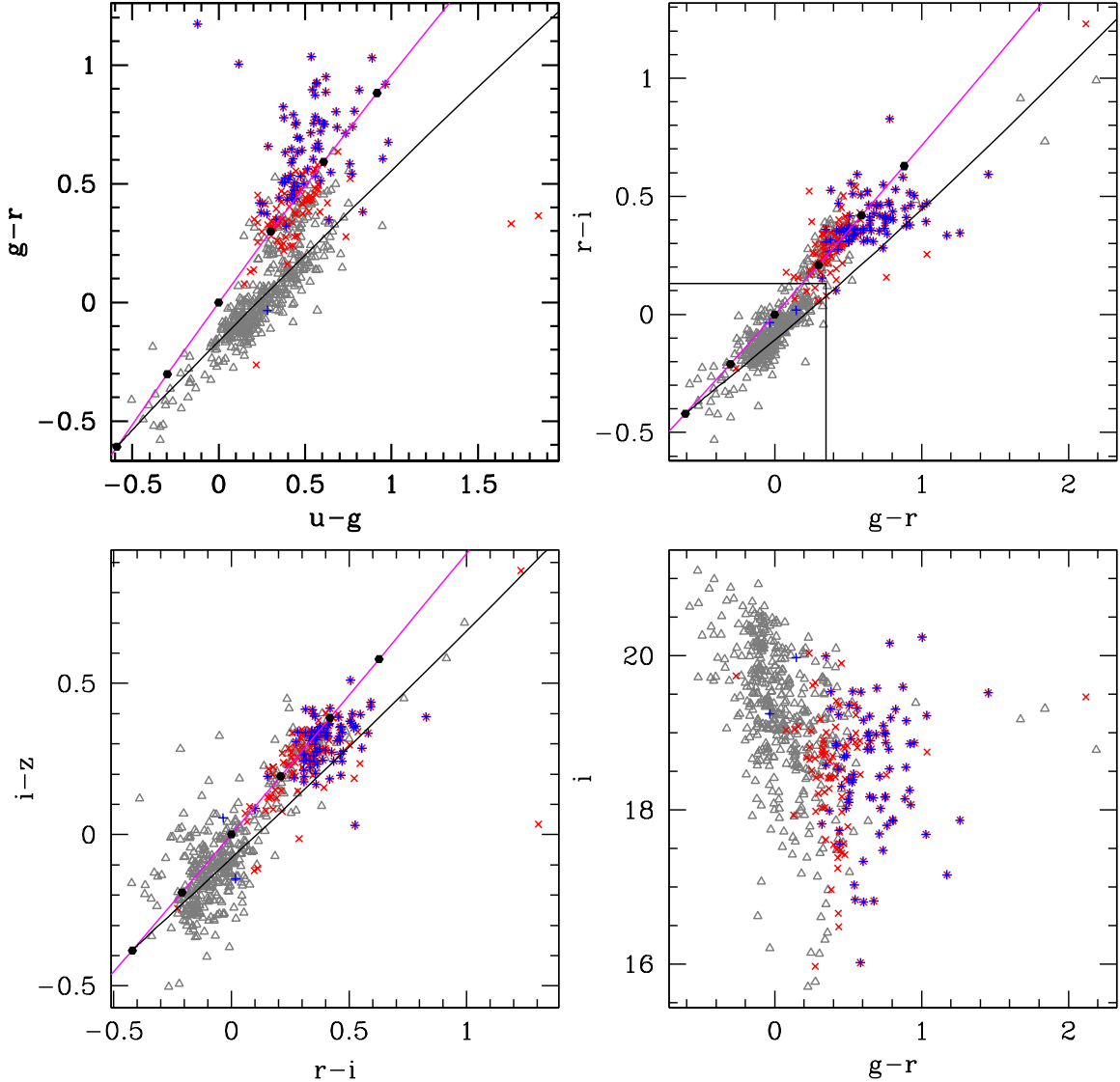}
\figcaption{Similar to Figure~\ref{fig:pmcolor}, but broken up schematically 
according to image morphology and redshift. Grey triangles are objects with 
point-like morphologies and no identifiable spectral features for obtaining 
redshifts. Blue `+' signs are resolved objects, and red `$\times$' marks are 
objects with measured redshifts large enough to be cosmological; in many cases 
the latter two types of symbol overlap, especially for the reddest objects 
(those with strong host galaxy components). The localization of objects 
with redshifts or resolved morphologies is consistent with the separation 
according to proper motion that is apparent in Figure~\ref{fig:pmcolor}.
\label{fig:rrcolor}}
\end{figure}

Figure~\ref{fig:rrcolor} shows the same color-color diagrams as 
Figure~\ref{fig:pmcolor}, 
but the different symbols highlight objects with resolved morphologies 
or convincing cosmological redshifts (either from the pipeline or determined 
manually), that is, groups of objects that should be 
exclusively extragalactic in nature. 
The near-total absence of resolved objects or objects with redshifts in the 
blue-$gri$ region is rather striking, but in fact unsurprising, 
considering that significant host-galaxy starlight contributions 
(which enable most of our redshift measurements and are responsible for the 
extended morphologies) tend to move objects toward the red.

Of the 347 QFOs in the blue-$gri$ region, only five have 
measurable redshifts and only two are reported to be 
resolved by the SDSS photometric pipeline. The five 
blue-$gri$ objects with redshifts share the following other properties: 
(1) redshift (uncertain for three out of five) determined from 
weak broad emission lines rather than host galaxy absorption features; 
(2) unresolved optical morphology; and 
(3) proper motion either unreliable (2 objects) or $\mu<6$~mas~yr$^{-1}$ 
(3 objects), well below our detection threshold. Two of these 
objects have FIRST counterparts, and one of these is also a RASS source. 
These are all very interesting objects, but the (possible) presence of 
broad emission lines, though weak, already hints that they may differ in a 
significant way from classical BL Lacs. 

The two blue-$gri$ objects with resolved 
morphologies lack FIRST, NVSS and RASS counterparts, and they lack 
redshifts. Visual inspection of the images reveals that they are 
marginally resolved. These are the only QFOs in the sample that are spatially 
resolved but lack redshifts; we do not regard this as convincing
evidence of an extragalactic nature.

%----------------------------------------------------------------------------%

\section {Cross-correlations with X-ray and radio catalogs}
\label{sec:cross-corr}

\subsection{ROSAT All Sky Survey (RASS)}
\label{sec:rosat}

We cross-correlated the full sample of quasi-featureless objects with the 
RASS bright and faint source catalogs (BSC, \citealt{voges99}; FSC, 
\citealt{voges00}). The RASS covers virtually the entire sky in the 0.1--2.4~keV 
range with a typical limiting sensitivity of $10^{-13}$~erg~cm$^{-2}$~s$^{-1}$ 
and positional accuracy of 10--30~arcsec, for point sources. 
Using a matching radius of one~arcmin, 
we find 65 matches in the BSC and 32 in the FSC. Based on a second search 
using a matching radius of five~arcmin and applying a constant background model, 
we estimate the total expected number of mismatches within one~arcmin to be 
significantly less than one for the BSC and approximately one for the FSC.

In order to convert the RASS count rates to approximate X-ray fluxes, we use the 
Portable Interactive Multi-Mission Simulator (PIMMS; \citealt{mukai93}). 
We assume that the intrinsic X-ray spectrum in the (observed frame) energy 
range 0.1--2.4~keV is well-described by a power-law with frequency index 
$\alpha = 1.25$ (i.e., $f_{\nu}\propto \nu^{-1.25}$), intermediate between 
typical values for low-redshift LBL and HBL objects (cf.\ \citealt{sambruna97}). 
We modify the power-law spectrum according to the Galactic 
neutral hydrogen column density $N_{\mathrm{H}}$ toward each object, obtained 
using the {\sc FTOOLS}\footnote{http://www.heasarc.gsfc.nasa.gov/ftools/} 
\citep{blackburn95} program {\sc NH}, deriving the column density values from the 
maps of \citet{stark92}. Adopting this absorbed power law as the input spectrum, 
we use PIMMS to convert from observed ROSAT PSPC count rate to flux density 
at 1~keV in the observed frame. 
For objects without counterparts in either the BSC or the FSC, we use 
the same procedure to place approximate upper limits on the 1~keV flux density, 
assuming count rates just below the FSC detection threshold (i.e., six~counts 
divided by the exposure time appropriate for the object's 
coordinates\footnote{Exposure map available at 
http://wave.xray.mpe.mpg.de/images/rosat/survey/rass\_bsc/}).

%After correcting optical magnitudes for the Galactic extinction derived from 
%the \citet{schlegel98} map, we calculate the X-ray to 
%optical slope $\alpha_{\mathrm{ox}}$ between 1~keV$=2.42\times 10^{17}$~Hz and 
%$\nu_{\mathrm{eff}}\approx 4.85\times 10^{14}$~Hz~$=6200$~\AA\ 
%(corresponding to the SDSS $r$ filter; 
%\citealt{ugrizsystem}) in the observed frame. This definition differs 
%somewhat from the standard convention, both in the fact that we do not 
%correct for redshift, which is unknown for many of our objects, and in our 
%choices of reference frequencies (see \S\ref{sec:alphas}). 

\subsection{FIRST survey and NVSS}
\label{sec:first_nvss}

We also cross-correlated the sample of QFOs against the Faint Images 
of the Radio Sky at Twenty~cm catalog (FIRST; \citealt{becker95,white97}) 
and the NRAO/VLA Sky Survey (NVSS; \citealt{condon98}). 
The FIRST survey is designed to cover approximately the SDSS footprint area, with 
a typical sensitivity of one~mJy at 1.4~GHz for point sources, a resolution of 
five~arcsec, and sub-arcsec positional accuracy. The NVSS covers the whole 
sky north of declination $-40^{\circ}$ with a resolution of 45~arcsec, rms 
positional accuracy better than seven~arcsec, and a typical limiting 
sensitivity of 2.5~mJy, also at 1.4~GHz.

We adopted a two~arcsec matching radius for FIRST and a 10~arcsec 
radius for NVSS; optical/radio matching is discussed in detail by 
\citet{ivezic02}. We find 183 and 198 matches in FIRST and NVSS 
respectively, within these radii. Of the 198 NVSS matches, 
32 are not covered by FIRST; all the remaining 166 are detected by FIRST. 
Applying the same type of simple 
background analysis described in \S\ref{sec:rosat}, the total expected number of 
mismatches using these matching radii is near zero for FIRST and 
approximately one for NVSS. Our choice of a 10~arcsec matching radius for NVSS 
is somewhat conservative; among our QFOs, there are two objects for which bright 
FIRST counterparts ($f_{\nu}>50$~mJy) exist within two~arcsec of the SDSS 
position, but the nearest NVSS counterparts are separated by $>10$~arcsec 
from the SDSS position (12.2 and 25.4 arcsec separations), 
due to overlap with other nearby bright radio sources. 
However, the potential for confusion between SDSS and NVSS sources increases 
rapidly for matching radii $>10$~arcsec, and using a matching radius of 
20 or 30~arcsec would result in numerous false matches.

In order to identify any likely FIRST counterparts with optical-to-radio 
separations greater than the matching radius of two~arcsec (such as 
wide-separation double-lobed sources),
we visually examined FIRST cutout images for all objects 
for which FIRST data were available at the time of writing. Only one additional 
likely FIRST match was discovered. The FIRST source has a 
somewhat unusual elongated radio morphology that 
approximately overlaps the SDSS position (J231952.82$-$011626.8) 
with a centroid separation of 3.2~arcsec; SDSS~J2319$-$0116 (which is extended 
and shows strong host galaxy features) is the only apparent optical source in 
a SDSS mosaic image within five~arcsec of the FIRST centroid position. 
There is also a RASS BSC source with a separation of four~arcsec (well within the 
positional error) from SDSS~J2319$-$0116. We consider SDSS~J2319$-$0116 to be the 
likely source of both the X-ray and radio emission. 

With very few exceptions, such as SDSS~J2319$-$0116, 
FIRST counterparts of SDSS quasi-featureless objects clearly display 
compact core-dominated morphologies to the limit of the 
resolution of the FIRST survey. There are several instances of objects showing 
extended structures such as halos or weak one-sided jets, 
but no apparent lobe-dominated cases. 

Where both FIRST and NVSS data exist, we use 
FIRST to take advantage of its significantly finer angular resolution. 
For all objects observed and not detected by FIRST, 
we use the FIRST non-detections to set 
upper limits on flux density at 1.4~GHz of 
$f_{\nu,\mathrm{lim}}=5\times f_{\nu,\mathrm{rms}}+0.25\mathrm{mJy}$, where 
$f_{\nu,\mathrm{rms}}$ is the noise level (typically $\approx 0.15$~mJy) 
appropriate to the object's coordinates 
and 0.25~mJy is a correction for the CLEAN bias (e.g., \citealt{becker95}).
For objects lacking FIRST data, we do not use NVSS non-detections to set upper 
limits due to the potential for confusion as discussed previously.

%We calculate the optical to radio slope parameter 
%$\alpha_{\mathrm{ro}}$ between 
%$\nu_{\mathrm{eff}}\approx 4.85\times 10^{14}$~Hz~$=6200$~\AA\ 
%($r$ band) and 1.4~GHz (FIRST and NVSS) in the observed frame, after correcting 
%the SDSS magnitudes for Galactic extinction as in the previous section. 
%Our rather unconventional use of 1.4~GHz as a reference frequency may 
%result in a systematic offset of our $\alpha_{\mathrm{ro}}$ values with 
%respect to $\alpha_{\mathrm{ro}}$ values published elsewhere; 
%see \S\ref{sec:alphas}.

%-------------------------------------------------------------------------------%

\section{Properties of BL Lac candidates}
\label{sec:multiwavelength}

Table~1 contains positions, SDSS photometric data, redshifts and proper motions 
for probable BL Lac candidates. Table~2 contains RASS, FIRST and NVSS 
data and derived quantities for probable BL Lac candidates. Tables~3 and~4 
contain the analogous information for the possible BL Lac candidates.

%-------------------------------------------------------------------------------%

\subsection{Redshift distribution}
\label{sec:zhist}

\begin{figure}
\epsscale{1.0}
\plotone{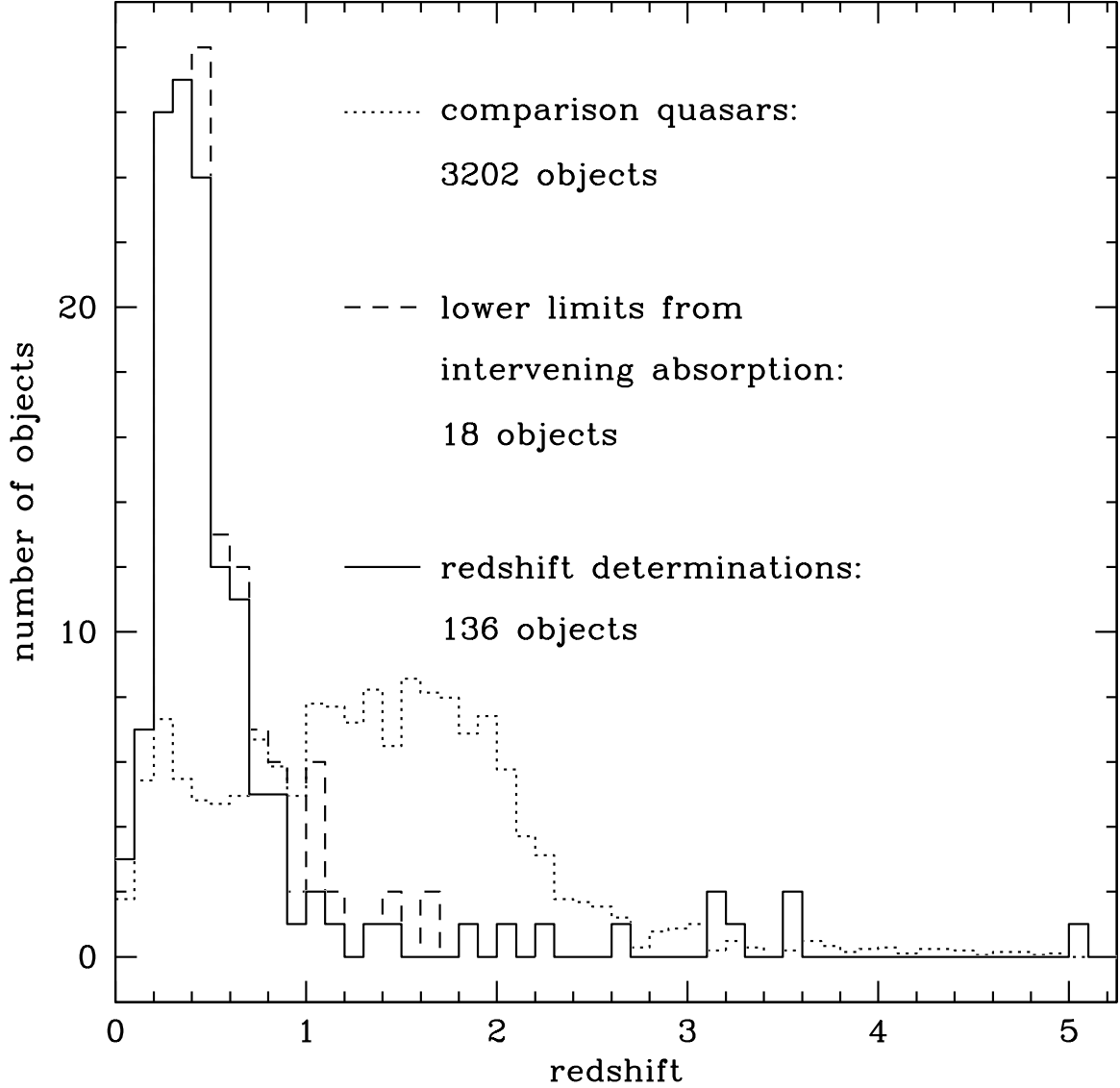}
\figcaption{Redshift distributions for BL Lac candidates with measurable 
redshifts, as well as the sample of comparison quasars from 
\S\ref{sec:pm_assess}. The solid line shows BL Lac candidates 
with emission-line or host-galaxy redshift determinations; the dashed line 
shows BL Lac candidates with redshift determinations plus those with 
redshift lower limits from intervening absorption. 
The dotted line shows the sample of comparison 
quasars (scaled down by a factor of approximately 20). 
The majority (232/386) of BL Lac candidates, including 86/240 probable 
candidates, are not shown due to the lack of redshift information.
\label{fig:zhist}}
\end{figure}

Figure~\ref{fig:zhist} shows the redshift distribution for the 154 
BL Lac candidates with measurable redshifts or lower limits from 
intervening absorption; by definition these all fall into the 
``probable'' category. The median measured redshift is 0.45. 
It is noteworthy that the redshift distribution of our BL Lac candidates 
extends out substantially farther than those for previous BL Lac 
samples (e.g., \citealt{stickel91,rector00,giommi05}).
A handful of objects extend to high enough redshift ($\mathbf{z}\ga 2.2$; 
$\mathbf{z}_{\mathrm{max}}=5.03$) that Ly$\alpha$ 
enters the observed wavelength range; these objects invariably show 
at least hints of weak, broad Ly$\alpha$ emission lines (typically with 
observed $EW>5$~\AA\ but $RFEW<5$~\AA), as in Figure~\ref{fig:line_examples}b. 
Some of these high redshift BL Lac candidates, along with similar objects 
discovered in high redshift quasar searches \citep{fan99,anderson01}, 
may in fact represent the low-EW tail of the normal quasar 
(high-luminosity AGN) population rather than true BL Lacs. 
All of the nine BL Lac candidates with $\mathbf{z}>2$ 
lack RASS counterparts, and six of them also lack FIRST/NVSS counterparts; 
thus it is difficult at present to confirm or reject this hypothesis 
(but see discussion in \S\ref{sec:alphas}). This possibility, in addition to 
the large sample size, may explain the apparent discrepancy in redshift 
distributions.

Of the BL Lac candidates with no redshift information, 
those that are truly extragalactic are likely to populate the 
redshift range $0.5\la \mathbf{z} \la 2.2$; lower than 
$\mathbf{z} \sim 0.5$, host galaxy features should generally be apparent even 
in moderate $S/N$ spectra, and at $\mathbf{z}>2.2$, the Ly$\alpha$ forest 
would enter the spectra. Also shown in Figure~\ref{fig:zhist} is the observed 
redshift distribution of the comparison quasar sample, which is close to 
flat out to $\mathbf{z}\sim 2$. Bearing in mind that a large fraction of the 
BL Lac candidates lacking redshifts are likely to be DC WDs, it would be 
impossible to make the BL Lac candidate redshift distribution match the 
comparison quasar redshift distribution up to $\mathbf{z}\sim 2$,
even with appropriate choices of redshift for the
BL Lac candidates with unknown redshifts or redshift limits.
This is qualitatively consistent with a picture in which BL Lacs are 
(on average) low-luminosity AGNs, and their
higher-luminosity counterparts, the FSRQs, do not necessarily 
share the property of having emission lines that are weak with respect to 
the beamed continuum radiation (e.g., \citealt{urry95}).

%-------------------------------------------------------------------------------%

\subsection{Optical colors of objects with X-ray and radio counterparts}
\label{sec:frcolor}

\begin{figure}
\epsscale{1.0}
\plotone{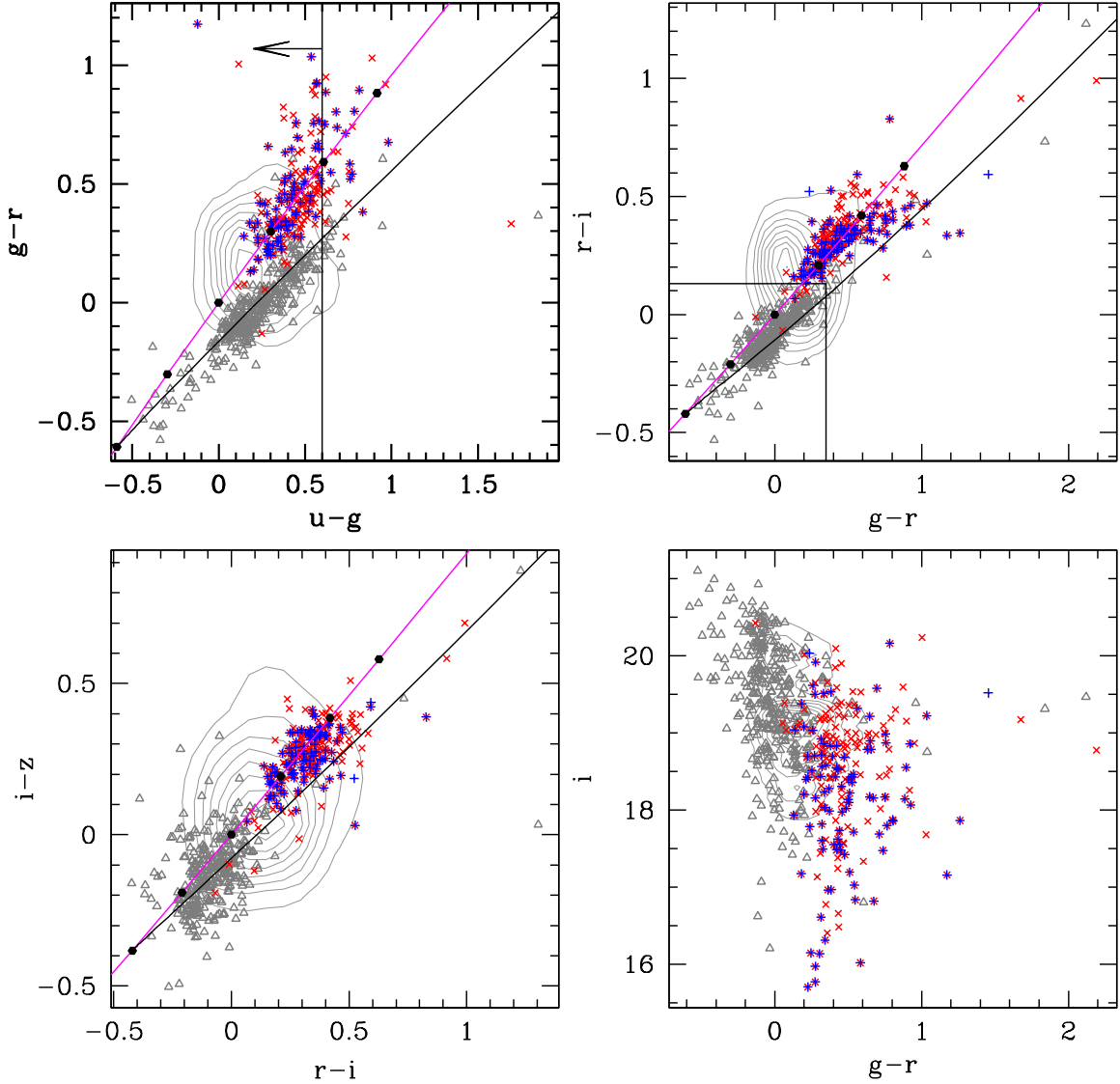}
\figcaption{Similar to Figure~\ref{fig:pmcolor}, but broken up schematically 
according to radio and X-ray detections rather than proper motion.
Contours represent the comparison quasar sample (see \S\ref{sec:pm_assess}). 
Grey triangles are QFOs with neither FIRST/NVSS nor RASS counterparts. 
Blue `+' symbols and red `$\times$' marks are QFOs with RASS or FIRST/NVSS 
detections, respectively (in many cases they overlap). The QFOs with X-ray 
and/or radio detections lie almost exclusively outside the region of color 
space populated by objects with significant proper motions 
(see Figure~\ref{fig:pmcolor}). 
\label{fig:frcolor}}
\end{figure}

Figure~\ref{fig:frcolor} shows the same color-color diagrams as 
Figure~\ref{fig:pmcolor}, symbolically coded according to X-ray and 
radio detections. Out of 97 RASS matches, only one 
lies in the blue-$gri$ region of color space. Out of 184 FIRST matches, only 
six lie in the blue-$gri$ region of color space. Out of 32 
objects with NVSS matches and no FIRST coverage, only one lies in the 
blue-$gri$ region. For comparison, 347/580$=59.8$\%
of all QFOs lie in the blue-$gri$ region of color space; after the 
proper motion cut, 156/386$=40.4$\% of BL Lac candidates still lie in this 
region. By contrast, there are only 22 QFOs \emph{outside} the blue-$gri$ 
region that \emph{do not} have FIRST or NVSS counterparts.
This simultaneously emphasizes that outside the blue-$gri$ region, there 
is not much room for contamination by DC WDs, and
reinforces the notion that true BL Lacs with 
such blue optical colors are much rarer than ``normal'' BL Lacs. 

\subsection{Subdividing the sample}
\label{sec:sample}

Primarily because we do not have reliable proper motion measurements 
for 246 QFOs, the sample of 386 BL Lac candidates remains 
significantly contaminated by DC white dwarfs. The observed color 
separation in Figures~\ref{fig:pmcolor}, \ref{fig:rrcolor} and 
\ref{fig:frcolor}, between objects with significant proper motions and 
objects with low proper motions or other evidence of an extragalactic/AGN 
nature (radio or X-ray detections or redshifts), argues strongly that 
most true BL Lac lie outside the blue-$gri$ region, while most DC WDs 
lie within it. Within the blue-$gri$ region there are a total of 10 
QFOs that show evidence of a likely extragalactic/AGN nature; 
outside the blue-$gri$ region there are a total of 11 QFOs that lack 
such evidence (plus an additional three rejected by the proper motion 
cut). Indeed, all available data are consistent with the 
more ambitious proposal put forward in \S\ref{sec:pmcolor}, that nearly 
all the QFOs within the blue-$gri$ region are stars whereas nearly 
all the QFOs outside the blue-$gri$ region are extragalactic.

Based on this compelling argument, and in order to minimize the impact 
of the remaining DC WD contamination, we subdivide 
the sample into ``probable'' BL Lac candidates and ``possible'' BL Lac 
candidates. Here ``probable'' signifies a probable extragalactic 
nature (though not necessarily classical BL Lac status) and ``possible'' 
signifies a likely stellar nature.
We define probable BL Lac candidates to be those objects 
that (1) lie outside the blue-$gri$ region of color space (i.e., $g-r\geq 0.35$ 
or $r-i\geq 0.13$) \emph{or} (2) have X-ray or radio 
counterparts or measured redshifts. We define possible BL Lac 
candidates to be everything else, that is, objects that lie within 
the blue-$gri$ region and have no indication of extragalactic 
nature (X-ray/radio counterparts or redshifts). 

By these definitions, we have 240 probable and 146 possible BL Lac 
candidates. Of the 39 \citet{kleinman04} objects that qualify 
as BL Lac candidates, 38 fall into the possible category; the 
remaining object also lies within the blue-$gri$ region, but has an 
apparent FIRST counterpart. Thus we conclude that the probable BL Lac 
candidates are not appreciably contaminated by previously known white dwarfs.
For the possible BL Lacs, we offer a word of caution: if there exist 
quasi-featureless extragalactic objects (no spectral features detected 
with the SDSS spectral $S/N$) that are weak compared to classical BL Lacs 
in X-ray and radio (undetected by RASS and FIRST/NVSS) and have 
bluer-than-average optical colors (within the blue-$gri$ region), they 
would fall into the possible category. Although we argue that most 
if not all of the possible BL Lac candidates are likely to be stars, 
a more detailed analysis of the statistics of DC WD proper motions and 
number counts would be required in order to prove that this is the 
case. Such an analysis is beyond the scope of the present work.

\subsection{BL Lac colors and number counts}

Adopting the conclusion that very few or none of the 
possible BL Lac candidates are in fact BL Lacs, we note that the colors 
of the true BL Lacs in our sample (the probable candidates) are 0.2--0.4~mag 
redder on average than the comparison quasars, shown as contours in 
Figure~\ref{fig:frcolor}. Starlight from 
BL Lac host galaxies can contribute to this effect, if the (apparent) 
AGN/host-galaxy luminosity ratio is lower for BL Lacs than for quasars, 
as expected if BL Lacs are on average relatively low luminosity AGNs. 
However, subtracting the host galaxy contributions cannot 
even approximately account for the differences in the average colors, so we 
conclude that the intrinsic power-law (host galaxy subtracted) spectra of our 
true BL Lacs are redder on average than the intrinsic (thermal/disk plus 
non-thermal) spectra of quasars. 

If we assume that the probable BL Lacs form a statistically complete sample 
to $i\approx 19$ (just brighter than the magnitude limit for low-redshift 
SDSS quasar spectroscopic target selection) and take into account that most 
of the probable candidates (all of those with $i<19$) have $g-i>0.2$, then 
the sample is complete to at least $g\approx 19.2$. (This may be a bad 
assumption because the sample may not be complete, or because not all objects 
in the sample may be true BL Lacs, but fortunately these two effects are of 
opposite sign.) There are 98 probable 
BL Lac candidates with $g<19.2$, selected out of 2860~deg$^{2}$ of sky, 
yielding a surface density of 0.034~deg$^{-2}$ brighter than this limit. 
If one is willing to make the additional assumption that $g\sim B$, this 
number may be compared with the number count extrapolations in Figure~2 
of \citet{padovani91}; they are mutually compatible, albeit with a 
large range of uncertainty. For comparison, there are $\sim 4$--5 quasars 
deg$^{-2}$ to the same limiting magnitude (e.g., \citealt{boyle00}).

%-------------------------------------------------------------------------------%

\subsection{SDSS target selection effects}
\label{sec:sdss_selection_effects}

The majority (349/386) of our BL Lac candidates, including 205/240 of the 
``probable'' candidates, have dereddened $u-g<0.6$ (marked in the upper left 
panel of Figure~\ref{fig:frcolor}), i.e., they satisfy the UVX criterion of SDSS 
low-redshift quasar target selection. Out of 386 total BL Lac candidates, 
225 are targeted by low-redshift quasar target selection, including 179/240 
probable candidates. The remainder of the candidates either have colors 
similar to WDs (and are therefore excluded from quasar target selection; 
\citealt{sdssqsotarget}) or are simply too faint.
The same UVX criterion provides grounds for inclusion in the blue-serendipity 
targeting routine as well; although this selection route 
receives lower priority than the main 
target selection routines, it populates parts of color-space that might 
otherwise be ignored (such as the WD exclusion region), 
and extends to a fainter limiting magnitude than low-redshift quasar 
targeting. Blue-serendipity targeting accounts for an additional 106 BL Lac 
candidates, including 10 probable ones. The remaining 55 
BL Lac candidates (including 51 probable ones) 
are mostly targeted due to their radio or X-ray properties.

%-------------------------------------------------------------------------------%

\subsection{Radio-optical and optical-X-ray slopes}
\label{sec:alphas}

For the purpose of calculating broad-band radio-to-optical and 
optical-to-X-ray spectral slopes ($\alpha_{\mathrm{ro}}$ and 
$\alpha_{\mathrm{ox}}$) that can be readily compared with those 
appearing in other works, we adopt standard (rest-frame) reference frequencies of 
5~GHz=6~cm, $6\times 10^{14}$~Hz=5000~\AA\ and $2.42\times 10^{17}$~Hz=1~keV.
Since we do not have observations covering these rest-frame frequencies 
in all cases (most notably near 5~GHz where we have essentially no data, 
the nearest observed frequency being 1.4~GHz from FIRST/NVSS), this procedure 
necessarily requires some assumptions and extrapolations. We assume that the 
radio, optical and X-ray spectra are locally power laws 
$f_{\nu}\propto \nu^{-\alpha}$ with exponents $\alpha_{\mathrm{r}}=-0.27$ 
(the average value for the 1~Jy sample; \citealt{stickel91}), 
$\alpha_{\mathrm{o}}=1.5$ (typical of the probable BL Lac candidates) 
and $\alpha_{\mathrm{x}}=1.25$ (see \S\ref{sec:rosat}). To objects that 
lack redshift information, we assign the median measured redshift (0.45) 
if they fall into the probable category, or zero if they fall into the 
possible category; uncertain redshifts and redshift lower limits are treated 
as secure redshift measurements for this calculation. We normalize the 
assumed radio and X-ray spectra using the 1.4~GHz and 1~keV (observed frame) 
data points. After correcting the optical magnitudes for the Galactic 
extinction derived from the \citet{schlegel98} map, we normalize the 
assumed optical spectrum using the nearest in wavelength to rest-frame 
5000~\AA\ of the $g,r,i,z$ magnitudes (depending on redshift). 

\begin{figure}
\epsscale{1.0}
\plotone{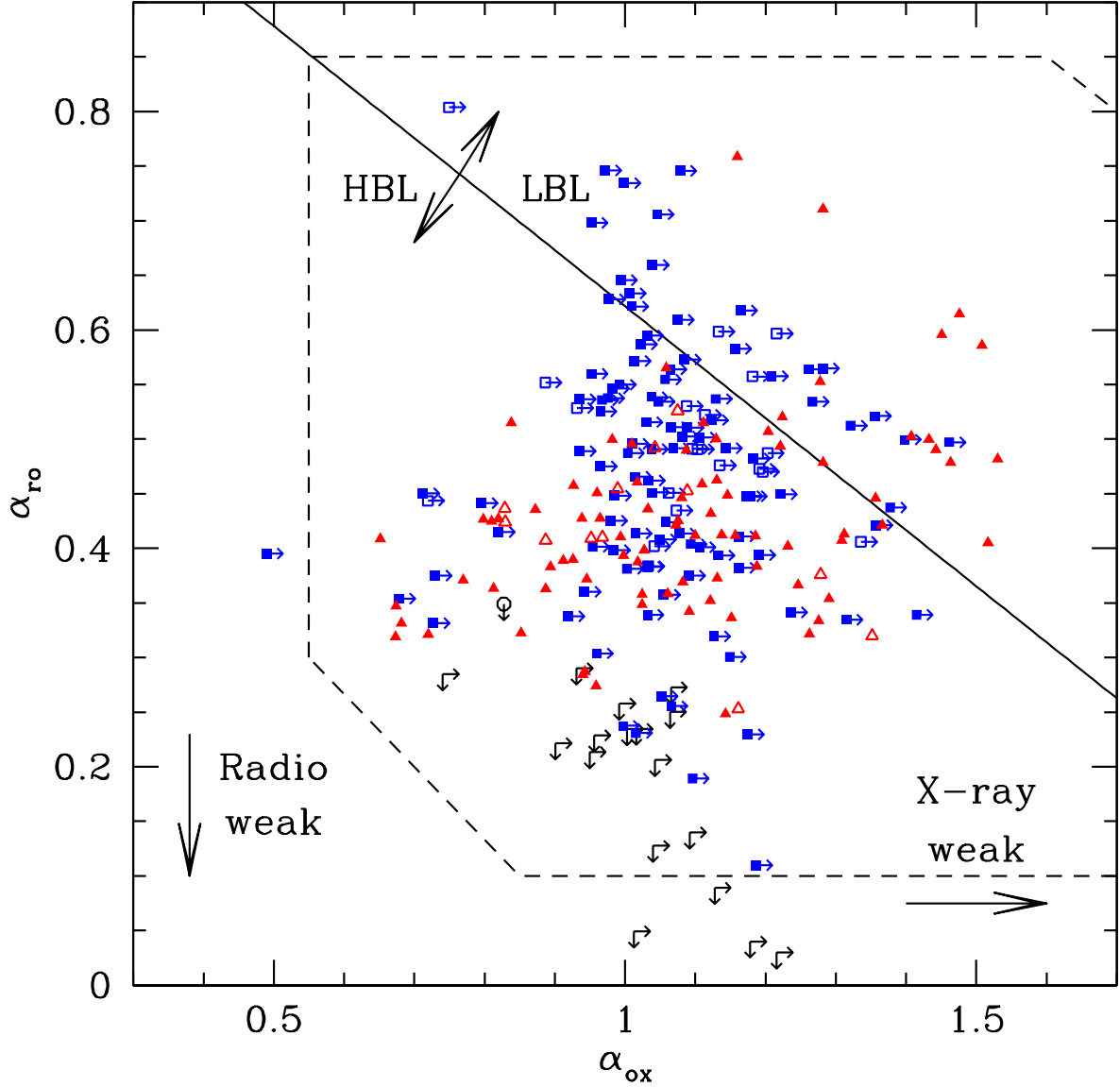}
\figcaption{\footnotesize
X-ray-optical-radio color-color diagram for probable BL Lac candidates. 
The parameters $\alpha_{\mathrm{ox}}$ and $\alpha_{\mathrm{ro}}$ are defined 
in the rest frame with the sign convention $f_{\nu}\propto \nu^{-\alpha}$, 
using reference points 
at 5~GHz, 5000~\AA\ and 1~keV as described in the text in \S\ref{sec:alphas}. 
The dashed line approximately encloses the population of known BL Lacs 
(e.g., Fig.~1 of \citealt{perlman01}). The slanted line shows the 
formal division between HBL-type and LBL-type objects 
at $\alpha_{\mathrm{rx}}\approx 0.75$ (e.g., \citealt{padovani95}).
Red triangles are objects with both radio (FIRST=solid or NVSS=outline) 
and X-ray (RASS) detections; blue squares with arrows are objects with radio
detections and X-ray upper limits. The black circle is the lone object with 
a RASS detection and a FIRST upper limit. Pairs of arrows represent 
objects with upper limits in both radio and X-rays. 
There are substantial uncertainties in the locations of objects on this 
diagram (see \S\ref{sec:alphas}). 
\label{fig:rxfig}}
\end{figure}

Figure~\ref{fig:rxfig} shows an X-ray-optical-radio color-color diagram 
($\alpha_{\mathrm{ox}},\alpha_{\mathrm{ro}}$)
for probable BL Lac candidates. There are uncertainties in the 
locations of individual objects on this diagram, mainly due to 
(1) non-simultaneity of observations at different frequencies, which is 
a concern because of BL Lac variability, (2) uncertainties introduced in the 
procedure of converting RASS count rates to flux, and (3) non-detection of 
many objects at X-ray or radio frequencies. Moreover, since we have 
assumed uniform X-ray, optical and radio power-law slopes to extrapolate 
observed data points to rest-frame reference frequencies, 
the true scatter in the plot is probably \emph{under}-represented.
We note that Figure~\ref{fig:rxfig} qualitatively differs from the analogous 
Fig.~3 of \citet{collinge04}, due to the more sophisticated procedure for 
calculating X-ray fluxes and broad-band spectral slopes; this simply serves to 
emphasize the potential for confusion.

It is nevertheless apparent that the sample follows a more-or-less 
continuous distribution of properties from high-energy peaked (HBL) toward 
low-energy peaked (LBL) SEDs, with no shortage of 
intermediate-type objects. This behavior is characteristic of the blazar class 
\citep{perlman98,perlman01}, despite previous and by now historical indications 
of a gap between HBL and LBL type objects (e.g., \citealt{padovani95}), which 
resulted from the very different selection biases imposed on radio versus X-ray 
selected samples. 
Perhaps the most noteworthy feature of Figure~\ref{fig:rxfig} is that all the 
objects with X-ray or radio detections fit so comfortably within the range of 
properties of previously-known BL Lacs (e.g., Fig.~1 of \citealt{perlman01}), 
despite the large sample size and relatively novel selection biases. 
If there are any BL Lac candidates in our sample with exceptionally weak 
radio or X-ray emission, then they must appear in Figure~\ref{fig:rxfig} 
only as limits or fall into the possible category (see discussion in 
\S\ref{sec:sample}). Among classical BL Lacs, the lowest known values of 
$\alpha_{\mathrm{ro}}$ are in the neighborhood of 0.1 (cf.\ \citealt{fan99}); 
four of our candidates have upper limits on $\alpha_{\mathrm{ro}}$ below this 
level. Because our derived $\alpha_{\mathrm{ro}}$ values are somewhat uncertain, 
we separate out all probable BL Lac candidates with $\alpha_{\mathrm{ro}}<0.2$ 
(8 objects) or no radio detections (another 19 objects). We present these 27 
candidates in Table~5; deeper radio observations of these objects are needed in 
order to understand the radio-weak tail of the BL Lac population.

In addition to the broadband spectral slope $\alpha_{\mathrm{ro}}$, 
another parameter that has been used to discriminate between radio-loud and 
radio-quiet AGN is the luminosity density at radio frequencies (e.g., 
\citealt{stocke92}); these authors placed the dividing line at 
$L_{\nu}\sim 10^{33}$~erg~s$^{-1}$~Hz$^{-1}$ for $\mathbf{z}\sim 2$ quasars. 
This number was obtained assuming an $H_0=50$~km~s$^{-1}$~Mpc$^{-1}$, 
$\Omega_{\mathrm{total}}=\Omega_{\mathrm{m}}=1$ cosmology, as opposed to 
the currently favored ``concordance'' $H_0=70$~km~s$^{-1}$~Mpc$^{-1}$, 
$\Omega_{\mathrm{m}}=0.3$, $\Omega_{\Lambda}=0.7$ model. Conveniently, 
these different cosmological models yield nearly equivalent (consistent to 
about 2\%) luminosity distance scales at $\mathbf{z}=2$ (the concordance 
model gives a slightly longer distance), so in practice no conversion is 
required. Assuming the concordance cosmology, 
all of the objects in Table~5 for which we have radio constraints 
fall below the approximate $L_{\nu}\sim 10^{33}$~erg~s$^{-1}$~Hz$^{-1}$ dividing 
line. While we do not advocate this approach to determining 
radio-loud/radio-quiet nature, this analysis provides another useful benchmark.

Bearing in mind the uncertainties in our derived values of $\alpha_{\mathrm{ox}}$
and $\alpha_{\mathrm{ro}}$, HBL or intermediate objects appear to constitute 
the majority of our sample. This may seem unsurprising, 
considering that the gap between the optical and soft X-ray bands is less than 
three orders of magnitude in frequency, while the gap between optical and radio 
is more like five. However, we note that the range of synchrotron peak 
frequencies $\nu_{\mathrm{peak}}$ of BL Lac objects is thought to be 
approximately $10^{13}\mathrm{Hz}<\nu_{\mathrm{peak}}<10^{18}\mathrm{Hz}$ 
(e.g., \citealt{anton04}), with LBLs at the low end and HBLs at the high end. 
Based on the experience of previous BL Lac searches --- 
``the surveys find what they can rather than what exists'' \citep{padovani01} --- 
we expected that our sample would be biased toward objects with 
$\nu_{\mathrm{peak}}$ close to the optical (i.e., toward the low end of the 
$\nu_{\mathrm{peak}}$ range), and thus dominated by LBLs. 
Uncertainties in ($\alpha_{\mathrm{ox}},\alpha_{\mathrm{ro}}$) 
notwithstanding, this is evidently not the case. 

A strong selection effect that may contribute to a HBL-bias is the SDSS 
quasar target selection UVX cut (dereddened $u-g<0.6$), which most of our 
BL Lac candidates satisfy, as discussed in \S\ref{sec:sdss_selection_effects}. 
Figure~\ref{fig:frcolor} shows that the numbers of BL Lac candidates drop 
sharply for $u-g>0.6$. Thus, most of our sample may be 
considered UVX-selected. Although the SDSS $u$-band is centered at 
$\nu_{\mathrm{eff}}\approx 8.51\times 10^{14}$~Hz (still toward the low end 
of the $\nu_{\mathrm{peak}}$ distribution), the requirement imposed on UV 
continuum slope may result in a bias toward HBL-type objects. 

%-------------------------------------------------------------------------------%

\subsection{Comparison with \citet{anderson03}}
\label{sec:anderson}

\citet{anderson03} presented a sample of candidate BL Lacs selected as 
part of a project to use SDSS imaging and spectroscopy to obtain optical 
identifications of a large sample of RASS X-ray sources. Their sample 
contained 38 ``probable'' (by their own definition) and seven 
additional ``possible'' (by their own definition) BL Lacs 
selected from 1400 deg$^2$ of SDSS data. This area of sky is covered 
by the spectroscopic parent sample for our search. Of the 38 \citet{anderson03} 
probable BL Lacs, our search recovers 35 objects; one of the remaining three 
objects fails the $S/N$ requirement (see \S\ref{sec:selection_sn}), and the other 
two fail the polynomial test (these objects are classified as galaxies; see 
\S\ref{sec:selection_stargal}). We also recover two out of the seven 
\citet{anderson03} possible BL Lacs. All the \citet{anderson03} 
objects that we recover are categorized (according to our definition) 
as probable BL Lacs due to their X-ray counterparts.

Following \citet{anderson03}, we calculate the X-ray to optical 
flux ratio $f_{\mathrm{x}}/f_{\mathrm{opt}}$, where $f_{\mathrm{x}}$ is the 
broadband absorption-corrected X-ray flux in the observed frame 
energy range 0.1--2.4~keV, and $f_{\mathrm{opt}}$ is the broadband 
extinction-corrected optical flux in the observed frame 
wavelength range 4000--9000~\AA. 
As in \citet{anderson03} we assume the 
X-ray and optical spectra to be power laws $f_{\nu}\propto \nu^{-\alpha}$ with 
$\alpha_{\mathrm{x}}=1.5$ and $\alpha_{\mathrm{opt}}=0.5$. 
The value of $\alpha_{\mathrm{x}}$ assumed here differs by 0.25 from that 
which we employed in \S\ref{sec:rosat}. The choice of 
$\alpha_{\mathrm{opt}}=0.5$ 
is representative of the sample of QFOs as a whole (see 
Figure~\ref{fig:pmcolor}), though they span a relatively large range ($\ga 3$) 
in $\alpha_{\mathrm{opt}}$; our probable BL Lac candidates are actually 
shifted toward steeper (redder) slopes as discussed in \S\ref{sec:pmcolor} 
and~\ref{sec:alphas}. We set the normalizations of the X-ray and optical 
power laws using the flux densities at 1~keV (cf.\ \S\ref{sec:rosat}) and
in the $g$-band ($\nu_{\mathrm{eff}}\approx 6.36\times 10^{14}$~Hz). 

\begin{figure}
\epsscale{1.0}
\plotone{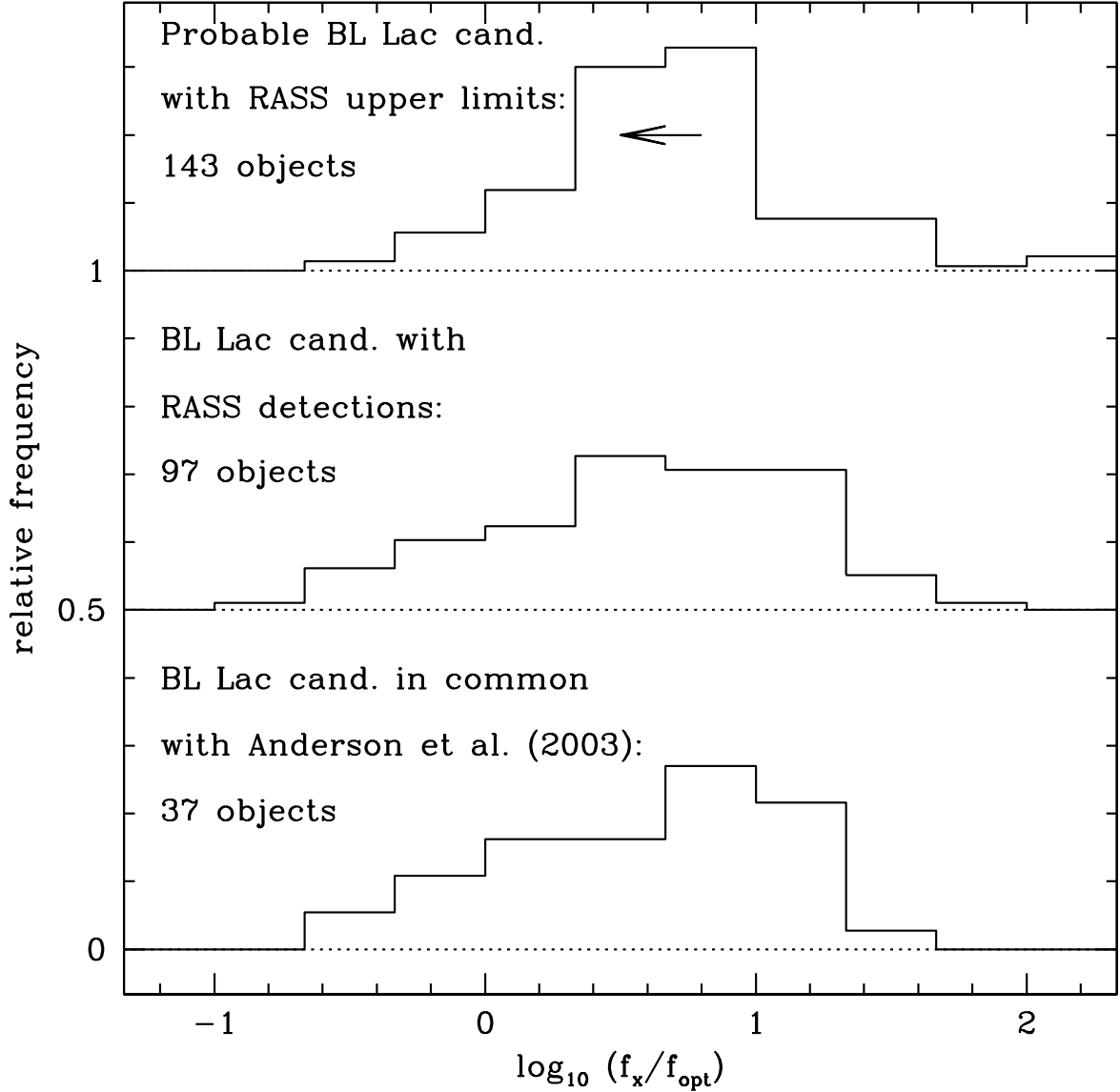}
\figcaption{
Histograms of $\log (f_{\mathrm{x}}/f_{\mathrm{opt}})$ for three 
subsets of BL Lac candidates. The bin size is 1/3 dex, and the middle and uppermost 
histograms are vertically offset for clarity by 0.5 and 1.0 respectively. 
The quantities $f_{\mathrm{x}}$ and $f_{\mathrm{opt}}$ are calculated as 
described in \S\ref{sec:anderson}. The lowermost histogram may be compared with 
the lower panel of Fig.~10 from \citet{anderson03}. 
The uppermost histogram consists of 
upper limits on $\log (f_{\mathrm{x}}/f_{\mathrm{opt}})$ (see \S\ref{sec:rosat}) 
for probable BL Lac candidates undetected in the RASS BSC and FSC.
\label{fig:fxfopt}}
\end{figure}

Distributions of $f_{\mathrm{x}}/f_{\mathrm{opt}}$ for several subsets of 
BL Lac candidates are shown in Figure~\ref{fig:fxfopt}. The lowermost 
histogram may be compared with the lower panel of Fig.~10 from \citet{anderson03} 
(not shown); 
the \citet{anderson03} distribution extends to a slightly higher value of 
$f_{\mathrm{x}}/f_{\mathrm{opt}}$. This probably reflects the different 
selection biases at work; if the eight \citet{anderson03} objects not included 
in our sample are fainter at optical wavelengths, 
they will tend to be biased toward X-ray dominated objects, for a fixed X-ray 
sensitivity. In any case, it is interesting that all three distributions in 
Figure~\ref{fig:fxfopt} and that from \citet{anderson03} 
peak at similar values of $f_{\mathrm{x}}/f_{\mathrm{opt}}$, suggesting a 
broad similarity in the properties of the X-ray-selected \citet{anderson03} 
sample and our (largely) optically-selected sample. This is consistent 
with the result from the previous section that the majority of our objects 
appear to be HBL-type.

%----------------------------------------------------------------------%

\section{Future work}
\label{sec:future}
The sample of BL Lac candidates presented in this paper is ripe for 
further analysis and future work. In particular:

\begin{itemize} 
\item None of the BL Lac candidates with radio or X-ray counterparts 
  stand out from the range of properties of previously known BL Lacs. 
  Our ability to constrain the 
  relative abundances of anomalously X-ray or radio-quiet BL Lacs is 
  currently limited by the sensitivity of available X-ray and radio 
  observations; thus deeper observations at these frequencies are 
  needed. 
\item With higher-$S/N$ spectroscopy of some of the objects that 
  currently lack redshift measurements, it would be possible to 
  construct the luminosity function of BL Lacs, and by comparing this
  with the luminosity functions of other radio-loud AGN, constrain models
  for jet opening angles as a function of luminosity. Carrying out
  this calculation will require a full understanding of the selection
  function of the sample imposed by the SDSS color selection 
  (\S\ref{sec:sdss_selection_effects}). For
  unreddened BL Lacs at $\mathbf{z} < 2$, this is reasonably straightforward,
  given the high completeness of the blue selection of SDSS quasars
  \citep{sdssqsotarget,vandenberk04b}.  
\item It will also be important to quantify the incompleteness of the
  sample due to our strict rejection of objects with emission
  lines. There should be BL Lacs whose host galaxies contain \ion{H}{2}
  regions, giving rise to emission lines of high enough EW to cause objects 
  to be rejected from the sample. There is also some inevitable 
  incompleteness at the low-luminosity end arising from our
  requirement that the non-thermal continuum be $>50\%$ of the 
  spectrum at either the blue or red end, as well as from our 
  requirements on $S/N$.  
\item One of the ways in which BL Lacs manifest themselves is by their
  variability. SDSS provides separate observations of these objects in the
  SDSS imaging and spectroscopy, which, as \citet{vandenberk04}
  show, can be used to explore the statistical properties of the
  variability of the population. Some patches of the 
  survey area have multiple imaging scans as well. It will be interesting 
  to see how our optically selected BL Lac candidates compare 
  in this respect with previously known BL Lacs.
\item Polarization is another defining characteristic of the BL Lac
  population. We have no information about the optical polarization
  properties of the sample, but the objects are reasonably bright, 
  with $i_{\mathrm{med}}=18.95$, so a
  broad-band polarization survey of a well-defined subset of the
  objects, especially those at higher redshifts, would be straightforward
  and informative.
\item Finally, the DC white dwarfs which we have culled
  from the sample are interesting in their own right \citep{kleinman04}.
  Their proper motions and brightnesses allow calculation of crude distances,
  and thus a determination of their
  luminosity function and a comparison with the population 
  of cooler white dwarfs. This may be important for models of white
  dwarf formation and cooling. 
\end{itemize}

%----------------------------------------------------------------------%

\section {Conclusions}
\label{sec:conclusion}

We present a large sample of candidate BL Lacertae objects from the 
spectroscopic database of the Sloan Digital Sky Survey. BL Lacs 
are rare AGNs in which the dominant source of observed radiation is thought 
to be a relativistic jet of plasma directed within a small angle to the line of 
sight, and in which the broad emission lines typical of unobscured AGNs are weak 
or washed out. While the latter trait is not well understood, the upshot is 
that BL Lacs have nearly featureless optical spectra; we use this 
property as the basis of our selection criteria. We define 
an object to have a quasi-featureless spectrum if it:
(1) has no identified emission lines with rest-frame equivalent width $RFEW>5$~\AA; 
(2) shows no evidence of stellar absorption features at zero redshift; 
(3) is not well-fit by stellar templates or combinations of galaxy eigenspectra;
and (4) has \ion{Ca}{2}~H\&K break strength $C\leq 0.4$, for $|C/\delta C|>3$.
(Certain extreme objects such as some broad-absorption-line quasars formally 
fulfill these requirements; such ``messy'' spectra are rejected as well.)

The primary astrophysical contaminants among quasi-featureless objects (QFOs) 
are DC white dwarfs. We have found (following \citealt{londish02}) that these 
can largely be rejected based on their high proper motions, although 
a significant fraction of objects lack reliable proper motion measurements, 
and hence a large degree of contamination remains in our sample. 
Out of 580 originally-selected QFOs, the 386 final BL Lac candidates 
are chosen to be those objects with low or unreliable proper motions.
Only 85 of the objects in the sample have been previously 
reported as BL Lacs in the NASA/IPAC Extragalactic 
Database\footnote{http://nedwww.ipac.caltech.edu/} (NED). 
The sample is drawn from approximately 2860 square degrees of sky ($\sim 1/3$ 
of the complete SDSS) and has a median $i$ magnitude of 18.95. 

We have been able to measure or constrain redshifts for 154/386 of our BL
Lac candidates, either from stellar absorption lines, or in a few
cases, by using intervening absorption systems to place lower limits.
The majority of these objects (131) have redshifts less than unity (the
median measured redshift is $\mathbf{z}_{\mathrm{med}}=0.45$), with 
a tail of objects extending out to high redshift 
($\mathbf{z}_{\mathrm{max}}=5.03$). These high redshift objects are 
potentially weak-lined normal quasars (as opposed to true BL Lacs), like 
those discussed by \citet{fan99} and \citet{anderson01} (but our sample 
does not include those specific objects due to the details of our 
selection criteria). The remaining true 
BL Lacs in the sample are likely to have redshifts in the approximate 
range $0.5\la \mathbf{z} \la 2.2$. 

We searched for X-ray and radio counterparts to QFOs using the 
ROSAT All-Sky Survey X-ray source catalogs and the FIRST and NVSS radio 
catalogs. On optical color-color diagrams, featureless objects with high 
proper motions tend to be confined in color space, generally following 
the loci of blackbody colors as a function of temperature (typically 
7000--12000~K). QFOs with redshifts or X-ray or radio counterparts 
(i.e., evidence of an extragalactic/AGN nature) 
almost exclusively lie outside the region of color space 
containing high-proper-motion objects. In fact, the optical colors of 
apparently-extragalactic QFOs are reasonably well-described 
in terms of power-law spectra (typically $1<\alpha_{\mathrm{opt}}<2$) 
plus host galaxy starlight rather than blackbodies. 
We conclude that most of the 156 remaining BL Lac candidates with 
blackbody-like colors are in fact DC white dwarfs, either with low proper 
motions or too faint to have reliable proper motion constraints.
To restrict the impact of this significant contamination on our 
results, we divide the sample into 240 ``probable'' (i.e., likely extragalactic) 
candidates, with colors unlike those of DC WDs or other evidence of 
extragalactic/AGN nature (redshift or radio/X-ray counterparts), 
and 146 ``possible'' (i.e., likely stellar) candidates, 
with colors consistent with the high proper motion objects, and no 
other evidence of extragalactic/AGN nature.
The probable BL Lac candidates have colors 
generally consistent with the ranges of known quasar colors, 
but on average are redder by $\sim 0.2$--0.4~mag, 
plus a redder tail of objects with more significant host galaxy 
contributions. 

Despite the large sample size and relatively novel selection biases, 
we find that the multi-wavelength properties of our probable BL Lac candidates 
do not systematically deviate from the range of properties of previously known 
BL Lacs. The majority of our BL Lac candidates detected in both X-ray and 
radio wavebands are consistent with being high-energy synchrotron-peaked 
(i.e., HBL/XBL type) objects, which may be due to a HBL-bias introduced 
by the UV-excess requirement ($u-g<0.6$) of SDSS low-redshift 
quasar spectroscopic target selection. 
Given the sensitivities of the RASS and FIRST/NVSS catalogs with which 
we compare, we find little evidence for a population of extragalactic
objects with featureless optical spectra, that are under-luminous in
either X-rays or radio. We present a list of the 27 most promising 
candidate radio-weak BL Lacs (though only eight currently 
have limits $\alpha_{\mathrm{ro}}<0.2$); 
more sensitive radio observations are needed 
for these objects before we can make a strong statement about the 
abundance of radio-quiet quasi-featureless AGN. 

%----------------------------------------------------------------------%
\begin{acknowledgements}

We are grateful to the anonymous referee for constructive suggestions.
MJC and MAS acknowledge the support of NSF grants AST-0071091 and AST-0307409.
MJC also acknowledges support from a National Defense Science and Engineering 
Graduate Fellowship. 

Funding for the creation and distribution of the SDSS Archive has
been provided by the Alfred P. Sloan Foundation, the Participating
Institutions, the National Aeronautics and Space Administration, the
National Science Foundation, the U.S. Department of Energy, the
Japanese Monbukagakusho, and the Max Planck Society. The SDSS Web site
is http://www.sdss.org/. 

The SDSS is managed by the Astrophysical Research Consortium (ARC)
for the Participating Institutions. The Participating Institutions are
The University of Chicago, Fermilab, the Institute for Advanced Study,
the Japan Participation Group, The Johns Hopkins University, the Korean
Scientist Group, Los Alamos National Laboratory, the
Max-Planck-Institute for Astronomy (MPIA), the Max-Planck-Institute
for Astrophysics (MPA), New Mexico State University, University of
Pittsburgh, Princeton University, the United States Naval Observatory,
and the University of Washington. 

\end{acknowledgements}

%----------------------------------------------------------------------%

%-------------------------------------------------------------------------
%FIGURES FIGURES FIGURES FIGURES FIGURES FIGURES FIGURES FIGURES FIGURES
%FIGURES FIGURES FIGURES FIGURES FIGURES FIGURES FIGURES FIGURES FIGURES
%FIGURES FIGURES FIGURES FIGURES FIGURES FIGURES FIGURES FIGURES FIGURES
%-------------------------------------------------------------------------

\clearpage

%-------------------------------------------------------------------------

\clearpage

%-------------------------------------------------------------------------

\clearpage

%-------------------------------------------------------------------------

\clearpage

%-------------------------------------------------------------------------

\clearpage

%-------------------------------------------------------------------------

\clearpage

%-------------------------------------------------------------------------

\clearpage

%-------------------------------------------------------------------------

\clearpage

%-------------------------------------------------------------------------

\clearpage

%-------------------------------------------------------------------------

\clearpage

%-------------------------------------------------------------------------

\clearpage

%-------------------------------------------------------------------------

%-------------------------------------------------------------------------

\begin{deluxetable}{cccccccccccc} 

\rotate 

\tabletypesize{\tiny} 

\tablewidth{0pt} 

\setlength{\tabcolsep}{0.05in}

\tablecaption{SDSS photometric and spectroscopic measurements, proper motions, 
and previous identifications for ``probable'' BL Lac candidates.}

\tablehead{ & $u$ & $g$ & $r$ & $i$ & $z$ & A$_r$~$^1$ & Optical & Target & & $\mu$ & Previously \\
IAU designation & (mag) & (mag) & (mag) & (mag) & (mag) & (mag) & morph. & code & Redshift & (mas yr$^{-1}$) & identified? \\
 (1) & (2) & (3) & (4) & (5) & (6) & (7) & (8) & (9) & (10) & (11) & (12)}

\startdata
 SDSS J012227.38$+$151023.1  &  20.34$\pm$0.05  &  19.90$\pm$0.02  &  19.55$\pm$0.03  &  19.25$\pm$0.03  &  18.93$\pm$0.05  &  0.17  &  p  &    1048580  &         \nodata  &    7.1  &        N  \\ 
 SDSS J012716.31$-$082128.9  &  20.26$\pm$0.07  &  19.70$\pm$0.03  &  19.14$\pm$0.03  &  18.81$\pm$0.02  &  18.55$\pm$0.04  &  0.08  &  p  &         20  &          0.362?  &    4.1  &        Y  \\ 
 SDSS J012750.83$-$001346.6  &  21.03$\pm$0.09  &  20.66$\pm$0.04  &  19.83$\pm$0.02  &  19.30$\pm$0.03  &  18.95$\pm$0.04  &  0.09  &  r  &    2097154  &          0.4376  &    6.0  &        N  \\ 
 SDSS J013408.95$+$003102.5  &  21.17$\pm$0.10  &  20.22$\pm$0.03  &  19.90$\pm$0.02  &  19.72$\pm$0.03  &  19.79$\pm$0.08  &  0.07  &  p  &          2  &         \nodata  &  154.5?  &        N  \\ 
 SDSS J014125.83$-$092843.7  &  18.11$\pm$0.02  &  17.59$\pm$0.03  &  17.21$\pm$0.02  &  16.96$\pm$0.01  &  16.68$\pm$0.02  &  0.08  &  p  &       7700  &     $\geq 0.5$?  &    5.8  &        Y  \\ 
 SDSS J020106.18$+$003400.2  &  18.90$\pm$0.02  &  18.62$\pm$0.02  &  18.25$\pm$0.03  &  17.99$\pm$0.02  &  17.80$\pm$0.02  &  0.07  &  r  &       5636  &          0.2985  &    7.6  &        Y  \\ 
 SDSS J020137.66$+$002535.1*  &  20.39$\pm$0.05  &  20.06$\pm$0.02  &  19.54$\pm$0.02  &  19.39$\pm$0.02  &  19.38$\pm$0.09  &  0.08  &  p  &    1048576  &         \nodata  &  176.1?  &        N  \\ 
 SDSS J022048.46$-$084250.4  &  18.92$\pm$0.02  &  18.62$\pm$0.03  &  18.27$\pm$0.02  &  18.03$\pm$0.02  &  17.80$\pm$0.03  &  0.06  &  p  &    1056276  &         0.5252?  &    7.1  &        N  \\ 
 SDSS J023813.68$-$092431.4  &  20.86$\pm$0.09  &  20.27$\pm$0.03  &  19.63$\pm$0.02  &  19.22$\pm$0.03  &  18.85$\pm$0.05  &  0.08  &  r  &    2102788  &          0.4188  &    3.6  &        Y  \\ 
 SDSS J024156.38$+$004351.6  &  20.66$\pm$0.08  &  20.03$\pm$0.02  &  19.61$\pm$0.02  &  19.41$\pm$0.03  &  19.29$\pm$0.05  &  0.09  &  p  &          3  &            0.99  &    6.1?  &        N  \\ 
\enddata

\tablecomments{Positions are in J2000 coordinates. Photometric measurements are 
reported in PSF magnitudes, not corrected for the Galactic extinction. The 
symbol `*' following the IAU designation indicates that the SDSS data using 
DR2 reductions were not available for the object; the data reported are from the 
DR1 reductions. 
The abbreviation `p' in column~(8) indicates that the object is point-like in the 
SDSS image; `r' indicates that it is resolved. Column~(9) reports the SDSS 
primary spectroscopic target code, in which each (binary) bit represents 
a flag for a different targeting algorithm; see \citet{sdssedr} for a list 
of the different targeting algorithms and their corresponding bits. 
A question mark `?' following the entry in column~(10) or column~(11) 
indicates that the redshift is uncertain or the proper motion is deemed 
to be unreliable, respectively. Column~(12) indicates whether 
the object was previously reported in NED as a BL Lac.
The full version of this table will be published online in the Astronomical 
Journal.}

\tablerefs{(1) Derived from \citet{schlegel98}} 

\end{deluxetable}

%-------------------------------------------------------------------------

\begin{deluxetable}{cc|cccccc|ccc|cc|cc} 

\rotate 

\tabletypesize{\tiny} 

\tablewidth{0pt} 

\setlength{\tabcolsep}{0.05in}

\tablecaption{Multi-wavelength data for ``probable'' BL Lac candidates.}

\tablehead{ & & \multicolumn{6}{c|}{ROSAT All-Sky Survey} & \multicolumn{3}{c|}{FIRST} & \multicolumn{2}{c|}{NVSS} & & \\
 IAU designation & $N_{\mathrm{H}}$ & match & sep. & count & exp. & $f_{\mathrm{E}}$(1keV)\tablenotemark{a} & $f_{\mathrm{x}}$\tablenotemark{b} & sep. & $f_{\nu}$(1.4GHz) & $f_{\mathrm{rms}}$ & sep. & $f_{\nu}$(1.4GHz) & $\alpha_{\mathrm{ox}}$ & $\alpha_{\mathrm{ro}}$ \\
 & ($10^{20}$cm$^{-2}$) & & (arcsec) & rate (s$^{-1}$) & (s) & & & (arcsec) & (mJy) & (mJy) & (arcsec) & (mJy) & & \\
 (1) & (2) & (3) & (4) & (5) & (6) & (7) & (8) & (9) & (10) & (11) & (12) & (13) & (14) & (15) }

\startdata
 SDSS J012227.38$+$151023.1  &    4.29  &  U  &  \nodata  &        $<$0.03  &   178.2  &  $<$5.7  &  $<$25.7  &  \nodata  &     \nodata  &  \nodata  &      0.8  &           38.0$\pm$1.2  &  $>$0.90  &     0.50  \\ 
 SDSS J012716.31$-$082128.9  &    3.97  &  U  &  \nodata  &        $<$0.01  &   438.8  &  $<$3.2  &  $<$14.4  &     0.30  &       172.2  &     0.20  &      0.8  &          111.9$\pm$3.4  &  $>$1.04  &     0.60  \\ 
 SDSS J012750.83$-$001346.6  &    3.23  &  U  &  \nodata  &        $<$0.01  &   432.2  &  $<$2.8  &  $<$12.3  &     0.30  &         5.6  &     0.15  &      4.2  &            7.3$\pm$0.5  &  $>$0.96  &     0.38  \\ 
 SDSS J013408.95$+$003102.5  &    2.91  &  U  &  \nodata  &        $<$0.02  &   373.9  &  $<$2.6  &  $<$11.3  &  \nodata  &      $<$1.1  &     0.17  &  \nodata  &                \nodata  &  $>$0.96  &  $<$0.25  \\ 
 SDSS J014125.83$-$092843.7  &    2.66  &  F  &  19.5  &  0.03$\pm$0.01  &   440.0  &     1.5  &      6.3  &     0.20  &       517.8  &     0.15  &      0.7  &         656.7$\pm$19.7  &     1.45  &     0.54  \\ 
 SDSS J020106.18$+$003400.2  &    2.67  &  B  &   1.9  &  0.34$\pm$0.03  &   400.0  &    17.0  &     72.1  &     0.40  &        13.9  &     0.14  &      2.0  &           12.7$\pm$0.6  &     0.90  &     0.34  \\ 
 SDSS J020137.66$+$002535.1  &    2.64  &  U  &  \nodata  &        $<$0.01  &   401.4  &  $<$2.5  &  $<$10.5  &  \nodata  &      $<$1.0  &     0.15  &  \nodata  &                \nodata  &  $>$1.02  &  $<$0.22  \\ 
 SDSS J022048.46$-$084250.4  &    3.05  &  B  &   9.5  &  0.09$\pm$0.02  &   315.0  &     4.9  &     21.2  &     0.20  &        57.3  &     0.13  &      0.9  &           64.9$\pm$2.4  &     1.10  &     0.45  \\ 
 SDSS J023813.68$-$092431.4  &    2.93  &  F  &  15.3  &  0.03$\pm$0.01  &   224.0  &     1.6  &      6.8  &     0.60  &         3.0  &     0.15  &  \nodata  &                \nodata  &     1.08  &     0.31  \\ 
 SDSS J024156.38$+$004351.6  &    3.45  &  U  &  \nodata  &        $<$0.03  &   226.6  &  $<$3.9  &  $<$17.1  &  \nodata  &      $<$1.0  &     0.14  &  \nodata  &                \nodata  &  $>$0.94  &  $<$0.22  \\ 
\enddata

\tablecomments{All frequencies refer to the observed frame. In column~(3), `B' 
indicates a match in the RASS bright source catalog; `F' indicates a match in 
the faint source catalog; `U' indicates the object was undetected, i.e., there 
was no match in either the BSC or the FSC within one~arcmin. Columns (7) and (8) 
are derived using slightly different assumed power-law slopes 
(see \S\S\ref{sec:alphas},\ref{sec:anderson}); upper limits on these quantities 
are obtained as described in \S\ref{sec:rosat}. Lack of data in columns~(10) 
and~(11) indicates that FIRST data were not available for the object. 
Column~(11) gives the FIRST rms flux level appropriate to the object's 
coordinates; this is used to set the upper limits in column~(10) as 
described in \S\ref{sec:first_nvss}. The full version of this table will be 
published online in the Astronomical Journal.}

\tablenotetext{a}{10$^{-13}$~erg~cm$^{-2}$~s$^{-1}$~keV$^{-1}$}
\tablenotetext{b}{10$^{-13}$~erg~cm$^{-2}$~s$^{-1}$}

\end{deluxetable}

%-------------------------------------------------------------------------

\begin{deluxetable}{cccccccccc} 

\tabletypesize{\tiny} 

\tablewidth{0pt} 

\setlength{\tabcolsep}{0.05in}

\tablecaption{SDSS photometric measurements and proper motions 
for ``possible'' BL Lac candidates.}

\tablehead{ & $u$ & $g$ & $r$ & $i$ & $z$ & A$_r$~$^1$ & Optical & Target & $\mu$\\
IAU designation & (mag) & (mag) & (mag) & (mag) & (mag) & (mag) & morph. & code & (mas yr$^{-1}$) \\
 (1) & (2) & (3) & (4) & (5) & (6) & (7) & (8) & (9) & (10)}

\startdata
 SDSS J015604.20$-$010029.0  &  19.27$\pm$0.03  &  18.76$\pm$0.03  &  18.58$\pm$0.02  &  18.54$\pm$0.02  &  18.65$\pm$0.03  &  0.08  &  p  &          4  &   95.3?  \\ 
 SDSS J020145.70$+$011013.0  &  20.97$\pm$0.28  &  20.14$\pm$0.04  &  19.99$\pm$0.03  &  19.97$\pm$0.04  &  20.12$\pm$0.10  &  0.07  &  r  &          2  &  \nodata  \\ 
 SDSS J020712.37$+$002443.2  &  20.28$\pm$0.06  &  20.15$\pm$0.03  &  20.23$\pm$0.03  &  20.37$\pm$0.04  &  20.51$\pm$0.16  &  0.08  &  p  &    1048576  &   37.3?  \\ 
 SDSS J021142.81$+$010446.3  &  18.46$\pm$0.02  &  18.83$\pm$0.02  &  19.35$\pm$0.02  &  19.72$\pm$0.03  &  19.89$\pm$0.09  &  0.08  &  p  &    9568256  &    0.0  \\ 
 SDSS J021925.07$+$004520.3  &  20.27$\pm$0.05  &  20.17$\pm$0.03  &  20.26$\pm$0.03  &  20.29$\pm$0.04  &  20.45$\pm$0.12  &  0.10  &  p  &    1048576  &    6.4?  \\ 
 SDSS J031303.31$-$061600.5  &  20.25$\pm$0.05  &  20.23$\pm$0.02  &  20.31$\pm$0.03  &  20.41$\pm$0.05  &  20.23$\pm$0.18  &  0.16  &  p  &   34603008  &    8.6?  \\ 
 SDSS J031729.73$-$071644.7  &  20.20$\pm$0.05  &  20.01$\pm$0.03  &  20.09$\pm$0.03  &  20.29$\pm$0.04  &  20.76$\pm$0.25  &  0.15  &  p  &  537919488  &  \nodata  \\ 
 SDSS J033444.87$-$011253.9  &  20.01$\pm$0.05  &  19.95$\pm$0.03  &  19.98$\pm$0.02  &  20.20$\pm$0.03  &  20.41$\pm$0.12  &  0.40  &  p  &    1048576  &   36.2?  \\ 
 SDSS J033556.39$-$002425.5*  &  20.73$\pm$0.07  &  20.31$\pm$0.02  &  20.11$\pm$0.02  &  20.03$\pm$0.03  &  19.97$\pm$0.12  &  0.34  &  p  &          2  &   19.0?  \\ 
 SDSS J034306.18$-$054806.3  &  18.48$\pm$0.02  &  18.36$\pm$0.01  &  18.48$\pm$0.01  &  18.63$\pm$0.02  &  18.83$\pm$0.04  &  0.14  &  p  &  538443776  &   13.6?  \\ 
\enddata

\tablecomments{Similar to Table~1, but for ``possible'' BL Lac candidates (see 
\S\ref{sec:sample}). The 
symbol `*' following the IAU designation indicates that the SDSS data using 
DR2 reductions were not available for the object; the data reported are from the 
DR1 reductions. The 
abbreviation `p' in column~(8) indicates that the object is point-like in the 
SDSS image; `r' indicates that it is resolved. Column~(9) reports the SDSS 
primary spectroscopic target code, in which each (binary) bit represents 
a flag for a different targeting algorithm; see \citet{sdssedr} for a list 
of the different targeting algorithms and their corresponding bits. 
A question mark `?' following the entry in column~(11) indicates that the 
proper motion is deemed to be unreliable. 
None of the possible BL Lac candidates have redshift 
measurements or previous BL Lac identifications in NED, so these columns have been 
excluded. The full version of this table will be published online in the 
Astronomical Journal.}

\tablerefs{(1) Derived from \citet{schlegel98}} 

\end{deluxetable}

%-------------------------------------------------------------------------

\begin{deluxetable}{cc|cccc|cc|cc} 

\tabletypesize{\tiny} 

\tablewidth{0pt} 

\setlength{\tabcolsep}{0.05in}

\tablecaption{Multi-wavelength limits for ``possible'' BL Lac candidates.}

\tablehead{ & & \multicolumn{4}{c|}{ROSAT All-Sky Survey} & \multicolumn{2}{c|}{FIRST} & & \\
 IAU designation & $N_{\mathrm{H}}$ & count & exp. & $f_{\mathrm{E}}$(1keV)\tablenotemark{a} & $f_{\mathrm{x}}$\tablenotemark{b} & $f_{\nu}$(1.4GHz) & $f_{\mathrm{rms}}$ & $\alpha_{\mathrm{ox}}$ & $\alpha_{\mathrm{ro}}$ \\
 & ($10^{20}$cm$^{-2}$) & rate (s$^{-1}$) & (s) & & & (mJy) & (mJy) & & \\
 (1) & (2) & (3) & (4) & (5) & (6) & (7) & (8) & (9) & (10)}

\startdata
 SDSS J015604.20$-$010029.0  &    2.57  &        $<$0.01  &   415.7  &  $<$2.4  &  $<$10.3  &      $<$1.0  &     0.14  &  $>$1.17  &  $<$0.15  \\ 
 SDSS J020145.70$+$011013.0  &    2.81  &        $<$0.01  &   410.2  &  $<$2.6  &  $<$11.0  &      $<$0.9  &     0.14  &  $>$0.95  &  $<$0.25  \\ 
 SDSS J020712.37$+$002443.2  &    2.74  &        $<$0.02  &   384.9  &  $<$2.5  &  $<$10.8  &      $<$1.0  &     0.15  &  $>$0.92  &  $<$0.27  \\ 
 SDSS J021142.81$+$010446.3  &    3.10  &        $<$0.02  &   303.5  &  $<$2.8  &  $<$11.9  &      $<$1.0  &     0.15  &  $>$1.04  &  $<$0.21  \\ 
 SDSS J021925.07$+$004520.3  &    3.17  &        $<$0.04  &   147.4  &  $<$5.7  &  $<$24.8  &      $<$0.9  &     0.13  &  $>$0.79  &  $<$0.27  \\ 
 SDSS J031303.31$-$061600.5  &    5.33  &        $<$0.03  &   226.6  &  $<$5.0  &  $<$23.5  &     \nodata  &  \nodata  &  $>$0.81  &  \nodata  \\ 
 SDSS J031729.73$-$071644.7  &    5.16  &        $<$0.02  &   271.6  &  $<$4.1  &  $<$19.2  &     \nodata  &  \nodata  &  $>$0.87  &  \nodata  \\ 
 SDSS J033444.87$-$011253.9  &    7.51  &        $<$0.01  &   536.7  &  $<$4.5  &  $<$21.9  &     \nodata  &  \nodata  &  $>$0.91  &  \nodata  \\ 
 SDSS J033556.39$-$002425.5  &    8.00  &        $<$0.01  &   526.8  &  $<$4.7  &  $<$22.7  &     \nodata  &  \nodata  &  $>$0.88  &  \nodata  \\ 
 SDSS J034306.18$-$054806.3  &    5.36  &        $<$0.01  &   409.1  &  $<$3.8  &  $<$17.9  &     \nodata  &  \nodata  &  $>$1.12  &  \nodata  \\ 
\enddata

\tablecomments{Similar to Table~2, but for ``possible'' BL Lac candidates (see 
\S\ref{sec:sample}). None of the possible BL Lac candidates are detected in 
the RASS, FIRST or NVSS catalogs, so RASS match codes and separations, FIRST 
separations, and NVSS separations and flux densities have all been excluded. 
The full version of this table will be published online in the Astronomical 
Journal.}

\tablenotetext{a}{10$^{-13}$~erg~cm$^{-2}$~s$^{-1}$~keV$^{-1}$}
\tablenotetext{b}{10$^{-13}$~erg~cm$^{-2}$~s$^{-1}$}

\end{deluxetable}

%-------------------------------------------------------------------------

\begin{deluxetable}{ccccccc} 

\tabletypesize{\tiny} 

\tablewidth{0pt} 

\setlength{\tabcolsep}{0.05in}

\tablecaption{Potential radio-weak BL Lac candidates.}

\tablehead{ & Optical & & $\mu$ & & & \\
IAU designation & morph. & Redshift & (mas yr$^{-1}$) & $\alpha_{\mathrm{ox}}$ & $\alpha_{\mathrm{ro}}$ & $\log_{10}{L_{5\mathrm{GHz}}}$\tablenotemark{a}\\
 (1) & (2) & (3) & (4) & (5) & (6) & (7)}

\startdata
 SDSS J004054.65$-$091526.8  &  p  &        5.03  &  \nodata  &     $>$1.01  &     $<$0.05  &     $<$32.6  \\ 
 SDSS J012155.87$-$102037.2  &  r  &      0.4695  &      8.1  &     $>$1.02  &     $<$0.23  &     $<$30.9  \\ 
 SDSS J013408.95$+$003102.5  &  p  &     \nodata  &   154.5?  &     $>$0.93  &     $<$0.29  &     $<$30.9  \\ 
 SDSS J020137.66$+$002535.1*  &  p  &     \nodata  &   176.1?  &     $>$0.99  &     $<$0.26  &     $<$30.8  \\ 
 SDSS J024156.38$+$004351.6  &  p  &        0.99  &     6.1?  &     $>$0.90  &     $<$0.22  &     $<$31.5  \\ 
 SDSS J024157.37$+$000944.1  &  p  &     0.7896?  &    57.8?  &        0.83  &     $<$0.35  &     $<$31.6  \\ 
 SDSS J025046.48$-$005449.0  &  p  &     \nodata  &    10.3?  &     $>$0.74  &     $<$0.28  &     $<$30.8  \\ 
 SDSS J025612.47$-$001057.8  &  r  &      0.6302  &  \nodata  &    $>$-0.11  &     $<$0.28  &     $<$31.1  \\ 
 SDSS J031712.23$-$075850.4  &  p  &      2.6993  &      4.5  &     $>$1.03  &     \nodata  &     \nodata  \\ 
 SDSS J090133.43$+$031412.5  &  r  &      0.4591  &      5.0  &     $>$1.04  &     $<$0.21  &     $<$30.8  \\ 
 SDSS J104833.57$+$620305.0  &  p  &     \nodata  &      5.0  &     $>$1.07  &     $<$0.27  &     $<$30.8  \\ 
 SDSS J114153.35$+$021924.4  &  p  &      3.5979  &      5.0  &     $>$1.19  &        0.11  &        32.9  \\ 
 SDSS J121221.56$+$534128.0  &  p  &        3.19  &      1.4  &     $>$1.22  &     $<$0.03  &     $<$32.3  \\ 
 SDSS J123743.09$+$630144.9  &  p  &      3.5347  &     6.4?  &     $>$1.18  &     $<$0.04  &     $<$32.4  \\ 
 SDSS J124225.39$+$642919.1  &  r  &      0.0424  &      5.1  &     $>$1.40  &     \nodata  &     \nodata  \\ 
 SDSS J133219.65$+$622715.9  &  p  &        3.15  &      3.2  &     $>$1.10  &        0.19  &        32.8  \\ 
 SDSS J142505.61$+$035336.2  &  p  &     2.2476?  &     3.6?  &     $>$1.13  &     $<$0.09  &     $<$32.1  \\ 
 SDSS J150818.97$+$563611.2  &  p  &     2.0521?  &      3.2  &     $>$0.95  &     $<$0.21  &     $<$32.0  \\ 
 SDSS J151115.49$+$563715.4  &  p  &     \nodata  &    10.6?  &     $>$1.06  &     $<$0.25  &     $<$30.8  \\ 
 SDSS J154515.78$+$003235.2  &  p  &     1.0114?  &      4.1  &     $>$1.00  &     $<$0.23  &     $<$31.9  \\ 
 SDSS J165806.77$+$611858.9  &  p  &  $\geq 1.41$?  &  \nodata  &     $>$1.09  &     $<$0.14  &     $<$31.7  \\ 
 SDSS J211552.88$+$000115.5  &  p  &     \nodata  &     4.5?  &     $>$0.95  &     \nodata  &     \nodata  \\ 
 SDSS J212019.13$-$075638.4  &  p  &     \nodata  &    28.7?  &     $>$0.96  &     $<$0.23  &     $<$30.8  \\ 
 SDSS J213950.32$+$104749.6  &  r  &       0.296  &     3.2?  &        1.02  &     \nodata  &     \nodata  \\ 
 SDSS J224749.55$+$134248.2  &  p  &     1.1746?  &      5.8  &     $>$1.10  &     \nodata  &     \nodata  \\ 
 SDSS J231000.81$-$000516.3  &  p  &  $\geq 1.68$?  &      5.0  &     $>$1.04  &     $<$0.13  &     $<$31.9  \\ 
 SDSS J232428.43$+$144324.4  &  p  &        1.41  &     2.2?  &     $>$0.75  &     \nodata  &     \nodata  \\ 
\enddata

\tablecomments{Selected data from Tables~1 and~2 for potential radio-weak 
BL Lac candidates (see \S\ref{sec:alphas}).}

\tablenotetext{a}{In units of erg~s$^{-1}$~Hz$^{-1}$, assuming $H_0=70$~km~s$^{-1}$, $\Omega_{\Lambda}=0.7$ and $\Omega_{\mathrm{m}}=0.3$. Luminosity density has been extrapolated to rest-frame 5~GHz as described in \S\ref{sec:alphas}. Uncertain redshifts and redshift limits have been assumed to be exact, and objects without redshift constraints have been placed at the median measured redshift of the sample ($\mathbf{z}_{\mathrm{med}}=0.45$).}

\end{deluxetable}

%-------------------------------------------------------------------------

\end {document}